\documentclass[10pt,twocolumn,letterpaper]{article}

\usepackage{template}
\usepackage{times}
\usepackage{epsfig}
\usepackage{graphicx}
\usepackage{amsmath}
\usepackage{amssymb}
\usepackage{xcolor}
\usepackage{subcaption}
\usepackage{adjustbox}
\usepackage{lipsum}
\usepackage{authblk}
\usepackage{cuted}

\usepackage{float}
\usepackage{makecell}

\usepackage{multicol}
\usepackage{algorithm}
\usepackage[noend]{algpseudocode}

\usepackage[aboveskip=0pt]{caption}
\newif\ifdrafting
\draftingtrue 
\ifdrafting
	\newcommand\blue[1]{\textcolor{blue}{#1}}
	\newcommand\celiu[1]{\textcolor{red}{Ce: #1}}
	\newcommand\huiwen[1]{\textcolor{red}{Huiwen: #1}}
	\newcommand\ds[1]{\textcolor{red}{Deqing: #1}}
\else
	\newcommand\blue[1]{}
	\newcommand\celiu[1]{}
	\newcommand\huiwen[1]{}
	\newcommand\ds[1]{}
\fi

\usepackage[pagebackref=true,breaklinks=true,letterpaper=true,colorlinks,bookmarks=false]{hyperref}

\usepackage[pagebackref=true,breaklinks=true,letterpaper=true,colorlinks,bookmarks=false]{hyperref}

\templatefinalcopy

\iftemplatefinal\fi

\begin{document}

\title{DVMark: A Deep Multiscale Framework for Video Watermarking}

\author[1]{Xiyang Luo}
\author[1]{Yinxiao Li}
\author[1]{Huiwen Chang}
\author[1]{Ce Liu}
\author[1]{Peyman Milanfar}
\author[1]{Feng Yang}
\affil[1]{Google Research}

\maketitle

\begin{strip}\centering
  \includegraphics[width=0.98\textwidth,trim=0 0 0 80pt]{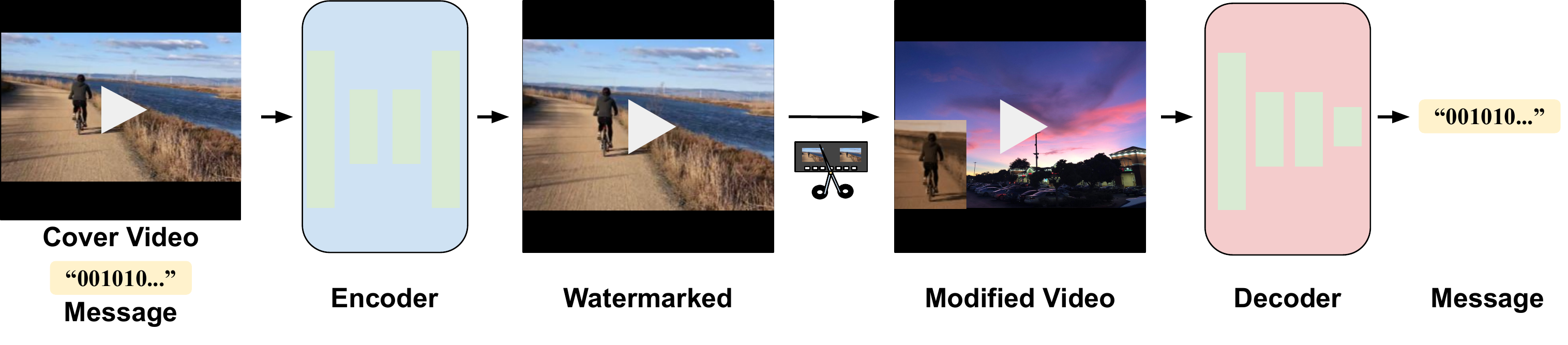}
  \vspace{10pt}
  \captionof{figure}{DVMark hides a message into a video which can be robustly recovered. The encoder network takes a cover video and a binary message, and generates a watermarked video that appears identical to the human eye as the input. The watermarked message can still be reliably extracted by our decoder even if the video undergoes a series of common video editing operations, such as compression, crop, color shift, and padding the watermarked video with other video content. }
  \label{fig:teaser}
\end{strip}

\begin{abstract}
\vspace{-10pt}
 Video watermarking embeds a message into a cover video in an imperceptible manner, which can  be retrieved even if the video undergoes certain modifications or distortions.  
 Traditional watermarking methods are often manually designed for particular types of distortions and thus cannot  simultaneously handle a broad spectrum of distortions. 
 To this end, we propose a  robust deep learning-based solution for video watermarking that is end-to-end trainable. 
 Our model consists of a novel multiscale design where the watermarks are distributed across multiple spatial-temporal scales. 
 It gains robustness against various distortions through a differentiable distortion layer, whereas non-differentiable distortions, such as popular video compression standards, are modeled by a differentiable proxy. 
 Extensive evaluations on a wide variety of distortions show that our method outperforms traditional video watermarking methods as well as deep image watermarking models by a large margin. 
 We further demonstrate the practicality of our method on a realistic video-editing application. 
\end{abstract}

\section{Introduction}
\label{sec:introduction}

\begin{figure*}
  \includegraphics[width=0.9\textwidth]{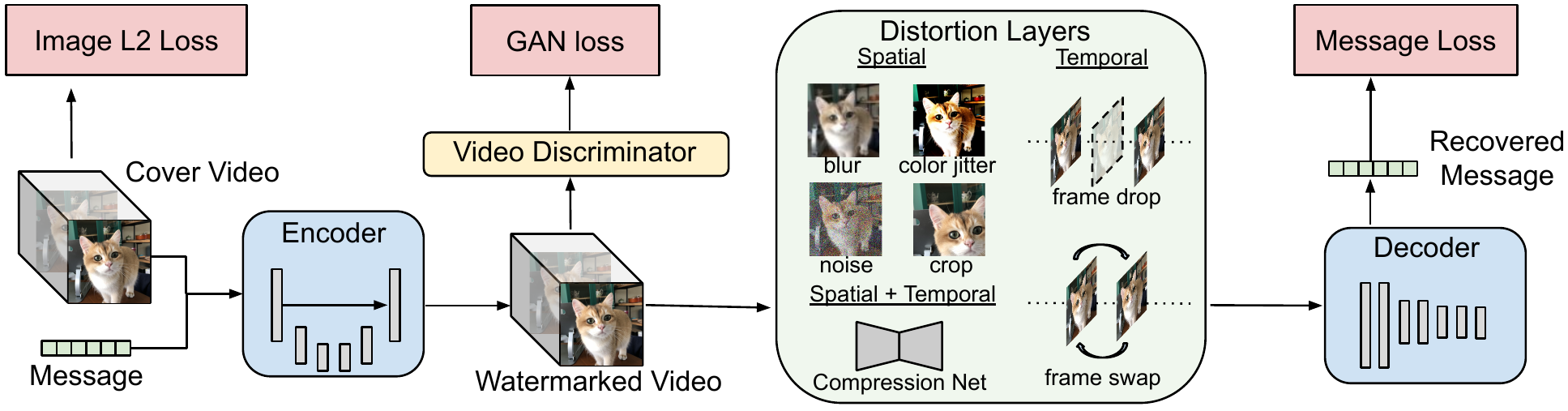}
  \centering
  \vspace{2mm}
  \caption{Our model consists of four main components, an encoder, a distortion layer, a decoder, and a video discriminator. The encoder takes as input a cover video and a binary message, and produces a watermarked video. The distortion layer applies a variety of distortions such as frame dropping, cropping, and compression to the watermarked video. The decoder network produces a predicted message from the distorted video. The video discriminator is trained to differentiate between the watermarked and cover videos, for the purpose of improving perceptual quality and temporal consistency.  }
  \vspace{-5pt}
  \label{fig:overall_architecture}
\end{figure*}

Watermarking is the problem of embedding messages in a cover media, such as images, video, or audio. Watermarks can be either visible or invisible. Invisible watermarks come with the benefit of being non-obtrusive to the original content, as well as making it more difficult to identify and detect for a potential attacker. In this paper we focus on the problem of invisible watermarks for videos, and refer to it as video watermarking throughout the rest of the paper.

There are three major factors to consider when evaluating a video watermarking system, namely, imperceptibility (or quality), robustness to various distortions, and number of message bits (payload or capacity). These factors are in tension with each other. For example, increasing robustness often comes at a price of lowering visual quality, and increasing the payload may have a non-trivial impact on both decoding accuracy as well as visual quality. The trade-offs between these factors determine the overall performance of a watermarking system. 

Traditional watermarking methods rely on hand-designed features to improve the system along these factors, such as applying various transforms or using perceptual masking~\cite{asikuzzaman2016robust, campisi2005video}. There are several drawbacks to a hand-crafted approach. First of all, different types of distortions often require different techniques, and therefore most traditional methods are not simultaneously robust to all types of distortions~\cite{asikuzzaman2017overview}. Secondly,  a hand-designed system often lacks the power to fully exploit the rich spatial-temporal information in a given video, leading to sub-optimal performance.
Recently deep learning based methods have been developed for image watermarking with promising results~\cite{zhu2018hidden,tancik2019stegastamp,luo2020distortion}. However, directly applying these deep image watermarking models to videos does not fully explore the temporal correlations in a video, resulting in a sub-optimal robustness-quality-payload trade-off (see Figure~\ref{fig:psnr_accuracy}, and Figure~\ref{fig:payload_accuracy}).

There are many challenges to designing a deep video watermarking method. For example, video compression is not  differentiable and hard to incorporate in an end-to-end training framework. Moreover, it is non-trivial to design a robust model that fully utilizes the temporal correlations in a video while retaining good temporal consistency and perceptual quality.  In this paper, we propose a highly robust deep video watermarking method that addresses all these issues. Below is a summary of our main contributions. 

\begin{itemize}
\vspace{-1.8mm}
    \item We propose DVMark, a highly robust deep learning based solution for video watermarking. Through rigorous evaluations on robustness, quality, capacity, as well as video distortions, we demonstrate that our method outperforms previous baselines by a significant margin. We further prove the practicality of our method by testing on multiple video resolutions and lengths, as well as on a video-editing application. 
    \vspace{-2mm}
    \item Our proposed method incorporates a novel multiscale design in both the encoder and decoder, where the messages are embedded across multiple spatial-temporal scales. This design brings more robustness compared to a single-scale network. 
    \vspace{-2mm}    
    \item We propose an effective solution for improving robustness towards video compression through the use of a differentiable proxy.
    \vspace{-2mm}
    \item Due to the prevalence of video editing, many videos may contain a mixture of watermarked and unwatermarked content. Therefore, we propose augmenting the decoder network with a watermark detector for detecting which frames are watermarked. This novel design gives the ability to precisely locate the watermarked frames.    
\end{itemize}

\section{Related Work}
\label{sec:related}

Watermarking digital media \cite{cox2002digital, katzenbeisser2000digital,bender1996techniques,jiansheng2009digital,o1997rotation,singh2013survey,tanaka1990embedding,hamidi2015blind} is an important and active area of research. Traditional video watermarking can be roughly classified into three groups: spatial domain, compressed domain, and transform domain methods, based on the domain it acts on~\cite{asikuzzaman2017overview}. 

Spatial domain methods embed the watermark by directly modifying the pixel values of the cover video. One example is Least Significant Bits (LSB)~\cite{kalker1999video,chopra2012lsb}, which is lacking in terms of robustness.  More advanced spatial domain methods such as block-based methods ~\cite{kimpan2004variable} or feature-point methods~\cite{kim2003invariant} addresses some of the weaknesses of LSB. Compression domain methods are designed to work for a specific video codec, such as the popular H.264~\cite{wiegand2003overview}. Zhang \emph{et al.}~\cite{zhang2007robust} proposed a method compatible with the H.264, where the watermark is embedded in the DCT coefficients. One drawback of this approach is that the watermark is format specific, and does not support conversion using an alternative codec. 

The third group of methods apply certain transforms to the cover video prior to embedding the watermark message. Popular transforms include Discrete Fourier Transform (DFT)~\cite{deguillaume1999robust}, and Discrete Cosine Transform (DCT)~\cite{dey2012dwt}, \emph{etc}. More complex transforms, such as the Discrete Wavelet Transform (DWT)~\cite{masoumi2013blind,chan2003dwt,liu2002robust,campisi2005video} and the Dual-tree Complex Wavelet Transform (DT-CWT)~\cite{coria2008video,asikuzzaman2016robust,agilandeeswari2016robust}, have also yielded good performance in terms of robustness and quality. In addition, techniques such as perceptual masking based on the Human Visual System (HSV)~\cite{swanson1998multiresolution,campisi2005video} further improve the perceptual quality for many traditional watermarking methods. A comprehensive survey for video watermarking could be found in \cite{asikuzzaman2017overview,patel2011survey}. 

Deep learning based approaches have recently been gaining traction in the field of watermarking, due to its impressive performance in terms of robustness and perceptual quality~\cite{zhu2018hidden,ahmadi2018redmark,wengrowski2019light,tancik2019stegastamp,wen2019romark}. HiDDeN~\cite{zhu2018hidden} was one of the first deep learning solutions for image watermarking. RedMark~\cite{ahmadi2018redmark} proposed extensions to the HiDDeN framework through the use of space-to-depth transforms and circular convolutions. Several works also focus on dealing with modelling more complex and realistic distortions~\cite{tancik2019stegastamp, wengrowski2019light, luo2020distortion}. StegaStamp~\cite{tancik2019stegastamp} models the camera capture distortion via a sequence of primitive transforms such as color-change, blurring, and perspective warping. Wengrowski \emph{et al.}~\cite{wengrowski2019light} proposed training a deep neural network to simulate screen display distortions. Liu~\emph{et al.}~\cite{liu2019novel} proposed a two stage pipeline for training robust models against style transfer distortions. Applying image watermarking methods to videos  results in a sub-optimal performance in terms of robustness and quality, since these methods are not able to utilize any temporal correlations between video frames.

For videos, VStegNet~\cite{mishra2019vstegnet} from Mishra \emph{et al.} achieved great performance for the task of video steganography. We note the difference with our work since VStegNet does not consider video distortions. The presence of even minor distortion greatly affects the quality and robustness of a model, and it is a non-trivial task to maintain both high perceptual quality and robustness. 

Generative Adversarial Network (GAN)-based training~\cite{goodfellow2014generative} is another important technique for deep learning based image and video processing for improving the perceptual quality of the generated videos. Techniques such as Wasserstein GANs~\cite{arjovsky2017wasserstein} and spectral normalization~\cite{miyato2018spectral} have greatly stabilized the training process. Some notable applications image and video processing where GANs were used include ESRGAN~\cite{wang2018esrgan}, TecoGAN~\cite{chu2020learning}, MOCOGAN~\cite{tulyakov2018mocogan}.

\section{Proposed Method}
\label{sec:method}

\begin{figure*}
  \centering
  \includegraphics[width=0.98\textwidth]{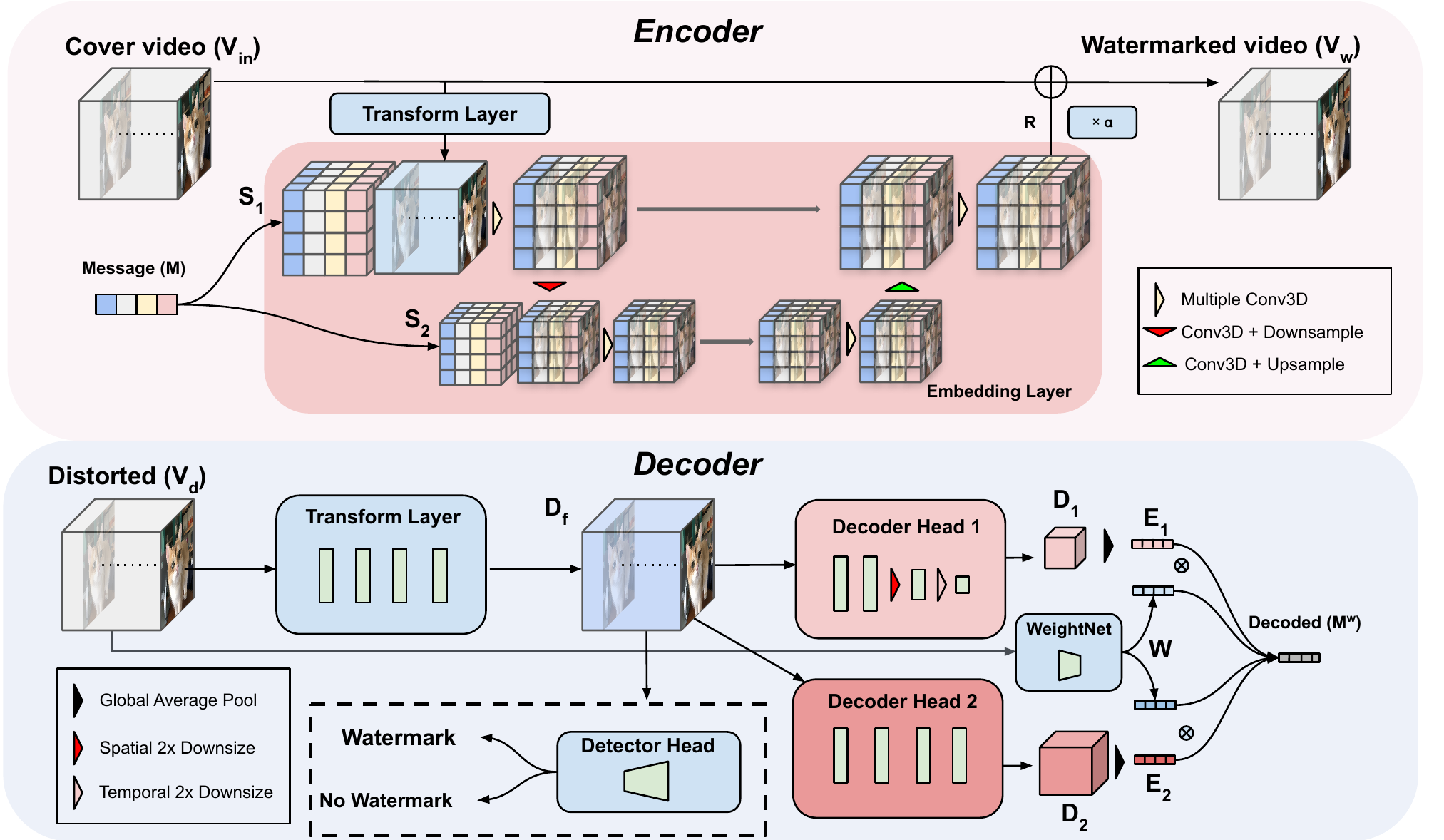}
  \vspace{2mm}
  \caption{Architecture of the multiscale encoder and decoder networks for DVMark. The encoder network fuses the input message with the cover video through a multiscale network on two different spatial-temporal scales. The input message is first repeated across spatial and temporal dimensions, and then fused with activations of the transformed cover video. The encoder learns a residual mapping which is added back to the cover video to obtain the watermarked video. A scalar factor could be multiplied to the residual to adjust the watermark strength at inference time. The decoder network consists of a transform layer and two detector heads which generates a decoded message map of different scales. The decoded maps are globally average pooled and linearly merged by a weight matrix of the same length predicted from a small network. The decoder is further augmented with a detector head with the ability to distinguish watermarked frames from unwatermarked frames. }
  \vspace {-13pt}
  \label{fig:encoder_architecture}
\end{figure*}

We propose an end-to-end trainable model for embedding a binary message of length $m$ to $n$ video frames. As shown in Figure~\ref{fig:overall_architecture}, our framework consists of four modules, an \emph{encoder}, a \emph{decoder}, a \emph{distortion layer}, and a \emph{video discriminator}. We explain in more detail each of the components below.

\subsection{Encoder}
The encoder network takes a cover video and an input message as input, and outputs the watermarked video. Figure~\ref{fig:encoder_architecture} (top) shows the architecture of the encoder network. The encoder consists of two main components: the \emph{transform layer} and the \emph{embedding layer}. The transform layer maps the input video sequence to a feature map of the same dimensions as the input. The embedding layer then outputs a watermark residual $R$. This is added to the original video multiplied by a scalar factor $\alpha$, ~\emph{i.e.}, 
\begin{equation}
    V_w = V_{in} +  R\times \alpha,
\end{equation}
where $V_{in}$ is the cover video, $V_{w}$ the watermarked video, $\alpha$ an adjustable watermark strength factor used to control the trade-off between quality and robustness.

\textbf{Transform Layer: }
The transform layer consists of 4 layers of 3D convolutions which transform the input video block to a 3D feature block with the same spatial-temporal dimensions. Each layer has 64 output channels, with stride equal to 1, spatial kernel size equal to 3, and temporal kernel size equal to 1, 1, 1, 3 respectively.  This design is inspired by traditional watermarking methods in the frequency-domain where the messages are embedded onto a transformed video instead of directly on the pixel domain. The transform layer thus allows us to learn an optimal transformation prior to merging with the messages. 

\textbf{Embedding Layer: }
The embedding layer (Figure~\ref{fig:encoder_architecture}) fuses the input message $M$ with the transformed video features at two scale levels $S_1$ and $S_2$. To merge the message with the video features, we first repeat the message along the spatial-temporal dimensions and concatenate with the feature map along the channel dimension. If the video feature map is a tensor of shape $t\times h\times w \times c$, the repeated message block has shape $t\times h\times w \times m$, with $m$ being the message length. For the first scale level $S_1$, We apply three Conv3D operations with kernel size 3 and number of channels 256, 128, 128 to the concatenated feature map. This operation is indicated by the yellow arrow labelled ``Multiple Conv3D" in Figure~\ref{fig:encoder_architecture}. For the second scale level $S_2$, we apply a $\text{AvgPool3D}$ to perform a 2X downsampling of the feature map from $S_1$ to $S_2$. We repeat the same process of concatenating with the message and applying Conv3D operations with channels 512, 256, 256 before upsampling through bilinear interpolation $S_1$. 


\subsection{Decoder}
The decoder network takes a possibly distorted version of the watermarked video $V_d$, and outputs a decoded message $M^w$. Similar to the encoder, our decoder also incorporates a multi-scale component through a multi-head design. 
We further predict a per-video weight vector for each decoder head, so that the distribution strategy across different scales can be content-adaptive. The weights are learned through a small network named WeightNet. 
Finally, we propose augmenting an additional head for the purpose of detecting whether a video has been watermarked. This has important practical implications since detecting whether a video (or parts of a video) has been watermarked is often as an important task as extracting the message itself. 

\textbf{Transform Layer: }
We first apply a transformation layer to map the distorted video $V_d$ to a feature map $D_f$ which will be shared by both the decoder and detector heads. We apply four Conv3D operations same as in the encoder transformation layer to extract the feature maps. In traditional frequency-based methods, the decoder usually applies the same frequency transforms as the encoder, but we allow the flexibility of learning different transformations for the decoder in our design.

\textbf{Multi-head Decoder: }
In our multi-head design, each decoder head outputs a decoded block $D_i$ with the same dimensions as each scale level $S_i$. For example with a training video size of $8\times128\times128\times3$,  $D_1\in \mathbb{R}^{4\times64\times64\times m}$ and $D_2\in \mathbb{R}^{8\times128\times128 \times m}$. Each decoder head contains four $\text{Conv3D}_{3,3,3}$ operations, where the output channels are 128,128,256,512 for head 1, and 128,128,128,256 for head 2. A 2X average pooling is applied to the spatial and temporal dimensions for the last two layers in head1. We then apply global average pooling to the blocks $D_i$ to obtain a decoded vector $E_i \in \mathbb{R}^m$, which represents the decoded information from each decoder head. Namely, 
\begin{equation}
    M^w = W^T E.
\end{equation}
The weight matrix $W_{ij}$ represents the importance of the predictions from scale $i$ for each bit $j$. $W$ is predicted per-video from a small network named WeightNet. \newline
\textbf{Watermark Detector: }
The watermark detector predicts whether a frame has been watermarked by the DVMark encoder or not. We observed that the feature maps learned from the transform layer can be re-used to differentiate between watermarked and unwatermarked video frames. Therefore, we train a detector head to differentiate watermarked frames from unwatermarked frames under the presence of various distortions.  The detector head consists of four $\text{Conv2D}_{3,3}$ operations, where the output channels are 128,128,256,512 respectively. 

\subsection{Distortion Layer}
Robustness to a broad variety of distortions has always  been a challenge for traditional video watermarking methods. In our framework, we achieve this by using a combination of common distortions during the training process. The distortions consist of temporal distortions, spatial distortions, as well as a differentiable version of video compression. At training time, each distortion is selected randomly with equal probability for each step of training. By randomly injecting distortions during the training process, both the encoder and decoder learn to be simultaneously robust to a variety of different distortions.

\textbf{CompressionNet: }
Robustness to video compression is a crucial requirement for a video watermarking method. Traditional video watermarking methods improve their robustness by embedding messages in the mid-low frequency components in a video. However, manipulations on the frequency domain alone cannot not fully capture the complexity of video codecs. To more accurately simulate video compression, we train a small 3D-CNN, named CompressionNet, to mimic the output of an H.264 codec at a fixed Constant Rate Factor (CRF)~\cite{wiegand2003overview}. With $\text{CRF}=25$, the final trained network gives a PSNR of 33.3dB with respect to the real compressed output. We then fix the weights of CompressionNet and apply it as a distortion when training the video watermarking model. 

\subsection{Video Discriminator}
\begin{figure}[!ht]
    \centering
    \includegraphics[ width=0.98\linewidth]{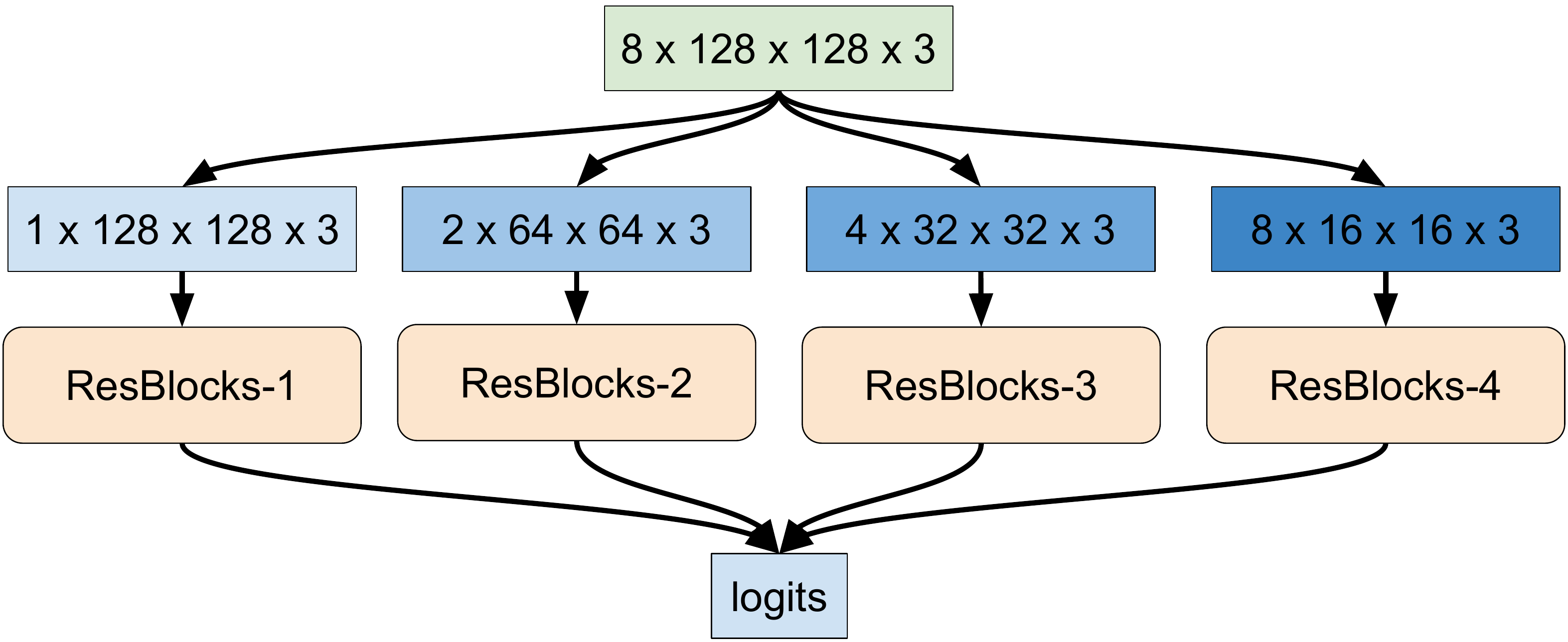}
    \vspace{2mm}
    \caption{Architecture of the multiscale video discriminator. Each 3D ResBlock processes an input of varying resolutions, allowing the network to capture both spatial and temporal differences between the cover and watermarked videos.}
    \label{fig:disc_res}
    \vspace{-3mm}
\end{figure}

Temporal consistency is an important issue for any video processing or generation tasks. Inspired by recent advances in video generations~\cite{saitotrain, tulyakov2018mocogan}, we tackle this issue by employing a multiscale video discriminator from TGAN2~\cite{saitotrain}. As shown in Figure~\ref{fig:disc_res}, the discriminator network consists of four 3D residual networks, each taking a different temporal and spatial resolution of the input video.

Given the discriminator $D$, the generator and discriminator losses are given by the Hinge loss formulation~\cite{kurach2018gan}. The discriminator optimizes $L_D$ while $L_G$ (see Eq.~\ref{eq:disc_gan_loss}) is added to the encoder-decoder loss in Eq.~\ref{eq:total_loss}.

\vspace{-5mm}

\begin{flalign}
 L_D &= \text{max}\{0, 1 - D(V_{w})\} + \text{max}\{0, 1 + D(V_{in})\}  \\ 
 L_G &= -D(V_{w}).
\label{eq:disc_gan_loss}
\end{flalign}
We empirically observe that a powerful video discriminator is crucial for achieving good temporal consistency for our video watermarking framework. GANs trained with only 2D convolutions in image space tend to exhibit more flickering compared to our models with 3D convolutions.

\subsection{Training}

\textbf{Encoder and Decoder: }
We first define the loss function which the encoder and decoder jointly optimize. Let $V_{in}$, $V_{w}$ denote the input and output of the encoder network, $M$ and $M^w$ denote the input and predicted message, we define the following objective function.

\begin{equation}
    L_{total} = L_I(V_{in}, V_{w}) + c_1L_M(M, M^w) + c_2 L_G(V_{w})
\label{eq:total_loss}
\end{equation}
where $c_i$ are the scalar weights for each loss term. Here $L_I$ is the pixel-wise $l_2$-loss. $L_M$ is the message loss given by the sigmoid cross-entropy. $L_G$ is the Wasserstein Hinge loss from the video discriminator, where we apply spectral normalization~\cite{miyato2018spectral} to the discriminator network. The encoder and decoder are trained jointly with respect to the loss above. 

\textbf{Watermark Detector:}
The detector is trained on top of a pre-trained encoder and decoder. For training, we randomly apply the DVMark encoder on each video clip with probability 0.5, and train the detector to predict whether the clip has been watermarked using a sigmoid cross entropy loss. Note that the weights of DVMark encoder and decoder are fixed during this process.

\begin{figure*}[!ht]
    \centering
    \begin{tabular}{c}
    \hspace{-0.25cm}
    \includegraphics[width=0.99\linewidth]{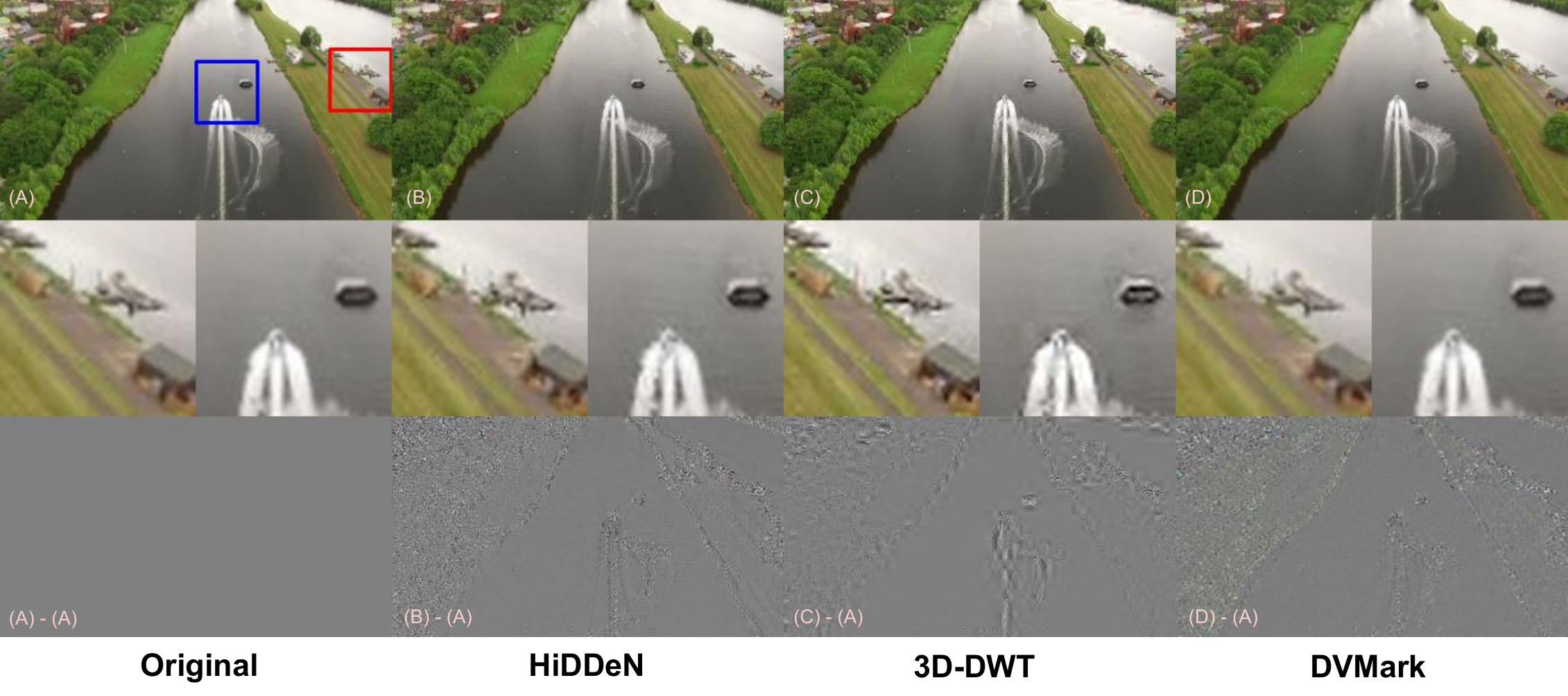}
    \end{tabular}
    \caption{Visual samples of original and watermarked video frames from various watermarking methods. From left to right: Original, HiDDeN, 3D-DWT, and DVMark. Top row contains the frames in full-view. Second row contains two zoomed-in patches for each image. The bottom row plots the normalized difference of the watermarked image with the original image, which provides a visualization of the watermarks from various methods.  Please enlarge and view the images on a computer screen. We see on the zoomed-in patches (second row) that DVMark induces less visible artifacts compared to the other methods.}
    \vspace{-5mm}
    \label{fig:visual_samples}
\end{figure*}


\section{Experiments}
\label{sec:experiments}

We train and evaluate our model on the Kinetics 600 dataset~\cite{carreira2018short, carreira2017quo}. This dataset is a large-scale video dataset which contains about 500K 10-second clips annotated with one of 600 human action categories~\cite{kay2017kinetics}.
About 390K clips are used for training, and the rest for evaluation.
We train our model on randomly cropped video clips with dimension $8\times128\times128$, and evaluate our model on 500 randomly selected videos in the Kinetics 600 validation set. We would like to emphasize that even though our model is trained on clips of size $8\times128\times128$, our model is applicable to videos of arbitrary dimensions and lengths, as indicated in Table~\ref{tab:accuracy-dimensions}.

\subsection{Comparison Methods}
We compare our model with both traditional transform-domain video watermarking methods, and a state-of-the-art deep image watermarking methods applied to videos.
 
\textbf{3D-DWT Method: } For the traditional method we choose a transform-domain method using 3D Discrete Wavelet Tranform (3D-DWT), a widely used technique for video watermarking method. We follow the implementation in~\cite{masoumi2013blind}, where a binary message is first transformed via the spread-spectrum technique~\cite{cox1997secure} and then added to the LLH, LHL, LHH, HLL, HLH, HHL sub-bands of a 3-level DWT. We refer to this method as 3D-DWT in the rest of the paper.

\textbf{HiDDeN~\cite{zhu2018hidden}: } 
We also compare our method with HiDDeN, a state-of-the-art deep image watermarking model robust to many image distortions, such as JPEG compression, crop and blur. For a fair comparison, we train HiDDeN with the same payload as DVMark, and apply the trained model with the same message to each video frame.   

\subsection{Method Evaluation}
In this section, we compare both the robustness and visual quality of our method against the two baseline methods.  To evaluate the robustness, we measure the bit accuracy, given by the percentage of bits correctly decoded. We report the bit accuracy of each model on a large collection of common distortions, including H.264, random frame drop, and frame averaging. All results in this section are evaluated on videos of size $128 \times 128 \times 8$, with payload $m=96$.

\begin{table*}[!ht]
    \begin{center}
    \begin{adjustbox}{width=0.99\textwidth}
    \Huge
    \begin{tabular}{cccccccccc|c} \hline
    & Identity & \makecell{H.264 \\ (CRF=22)} &\makecell{Frame Average \\(N=3)} &\makecell{Frame Drop \\ (p=0.5)} &\makecell{Frame Swap \\ (p=0.5)} &\makecell{Gaussian Blur \\(2.0)} & \makecell{Gaussian Noise \\(0.04)} &\makecell{Random Crop \\ (0.4)} &\makecell{Random Hue \\ (1.0)}  &\makecell{\textbf{Average}} \\ \hline
    \makecell{HiDDeN} &99.10  & \makecell{79.85}& \makecell{96.91} & \makecell{\textbf{99.03}}& \makecell{99.10} & \makecell{72.70} & \makecell{91.27} & \makecell{95.27}& \makecell{98.98} & \makecell{92.47}\\ \hline
    \makecell{3D-DWT} & 99.69 & \makecell{89.29} & \makecell{96.78}& \makecell{79.85} & \makecell{90.15} &\makecell{92.78} &  \makecell{\textbf{99.01}}& \makecell{71.38} & \makecell{\textbf{99.59}} & \makecell{90.95}\\ \hline
    DVMark & \textbf{99.99} & \makecell{\textbf{92.94}} & \makecell{\textbf{98.10}}& \makecell{98.99} & \makecell{\textbf{99.35}}& \makecell{\textbf{98.09}} & \makecell{98.56} & \makecell{\textbf{97.06}} & \makecell{\textbf{99.81}}& \makecell{\textbf{98.10}} \\ \hline    
    \end{tabular}
    \end{adjustbox}
    \end{center}
    \caption{\small{Comparison of Bit Accuracy for different watermarking methods. From top to down, we present the bit accuracy evaluated on 500 random videos for HiDDeN, 3D-DWT, and DVMark (our method), on a variety of common distortions. }}
    \label{tab:bit-error-rate}
\end{table*}

\begin{figure*}[!ht]
    \centering
    \begin{tabular}{c c c}
    \includegraphics[ width=0.32\linewidth]{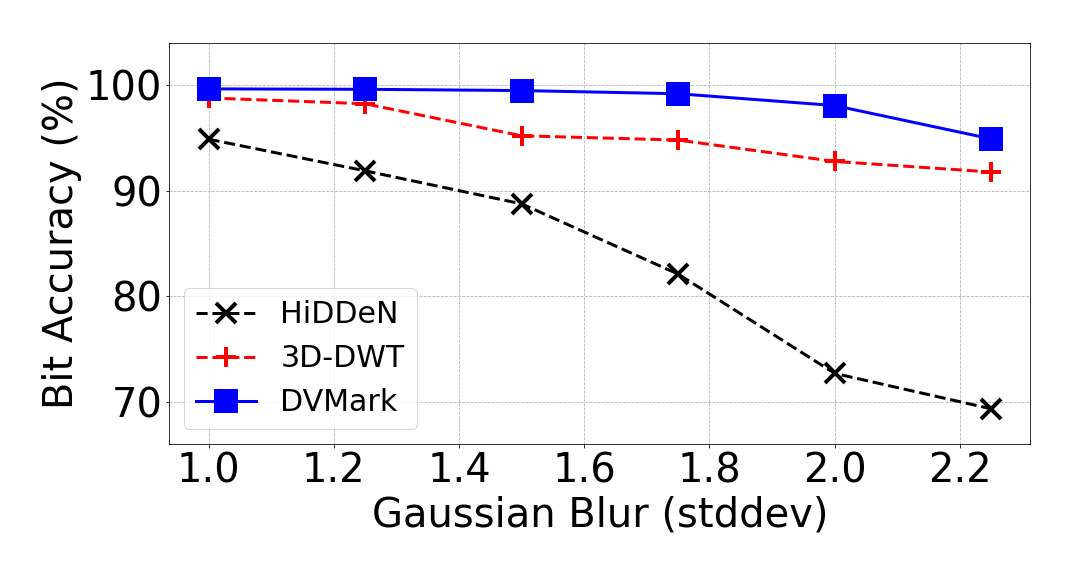} & \hspace{-3mm}
    \includegraphics[ width=0.32\linewidth]{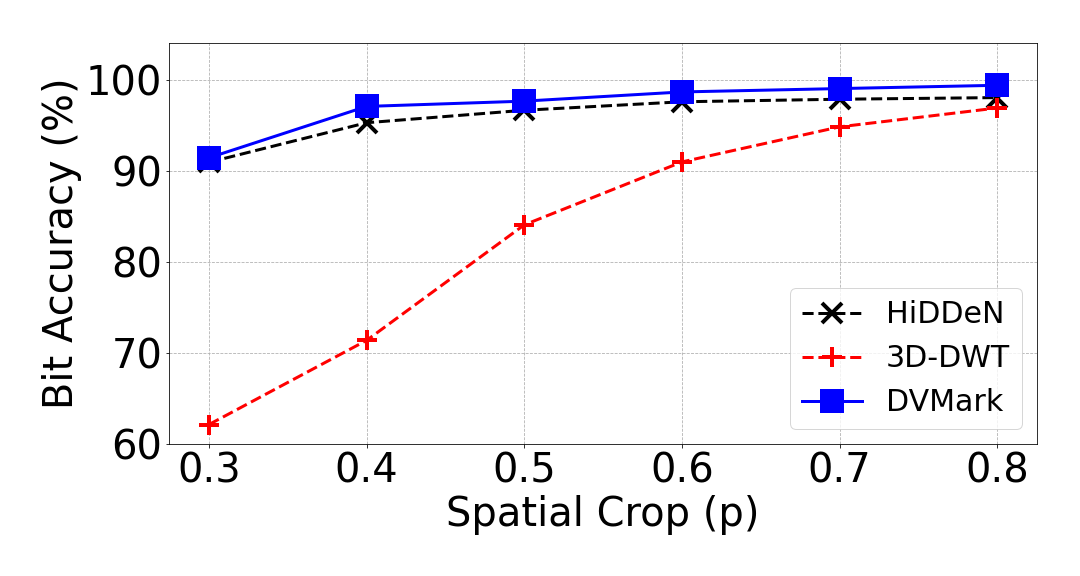} & \hspace{-3mm}
    \includegraphics[ width=0.32\linewidth]{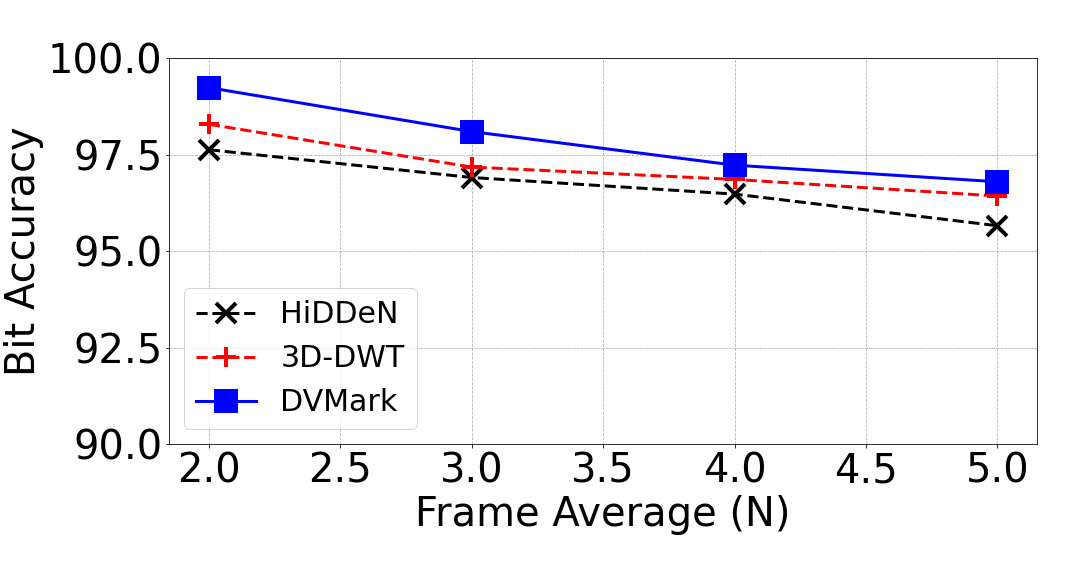} \\ 
    \includegraphics[ width=0.32\linewidth]{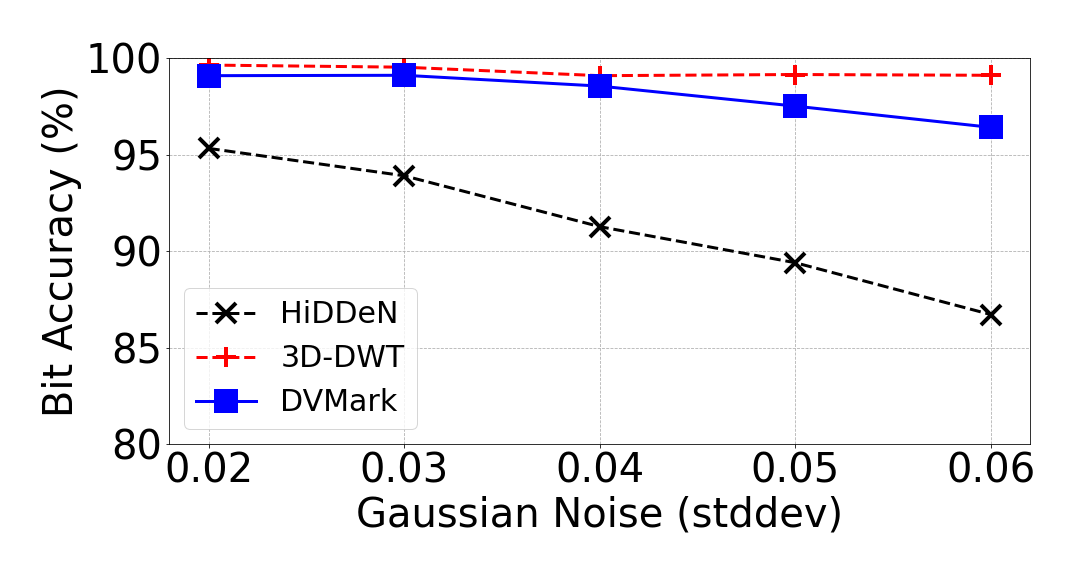} & \hspace{-3mm}
    \includegraphics[ width=0.32\linewidth]{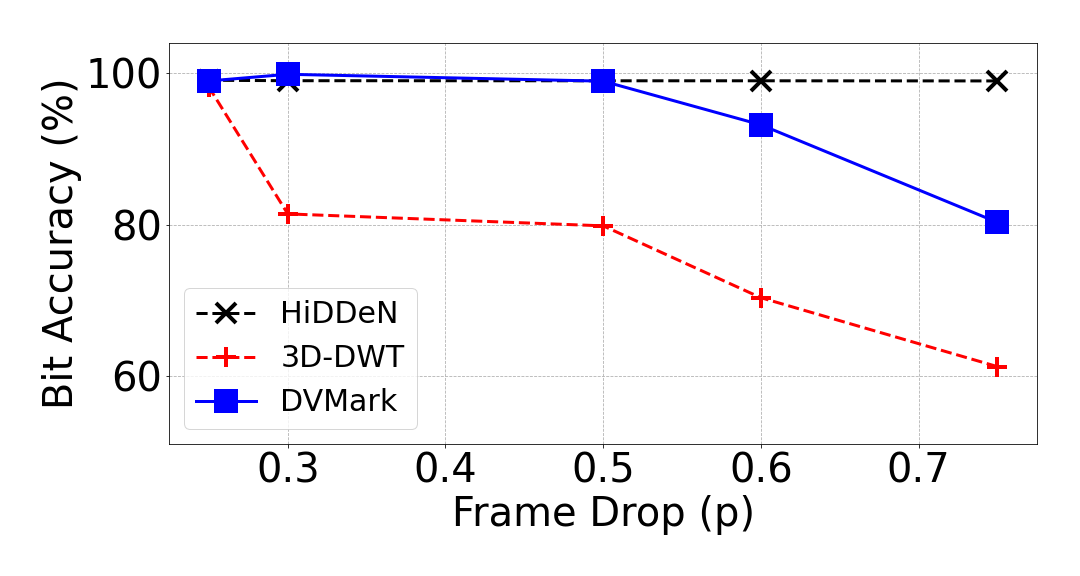} & \hspace{-3mm} 
    \includegraphics[ width=0.32\linewidth]{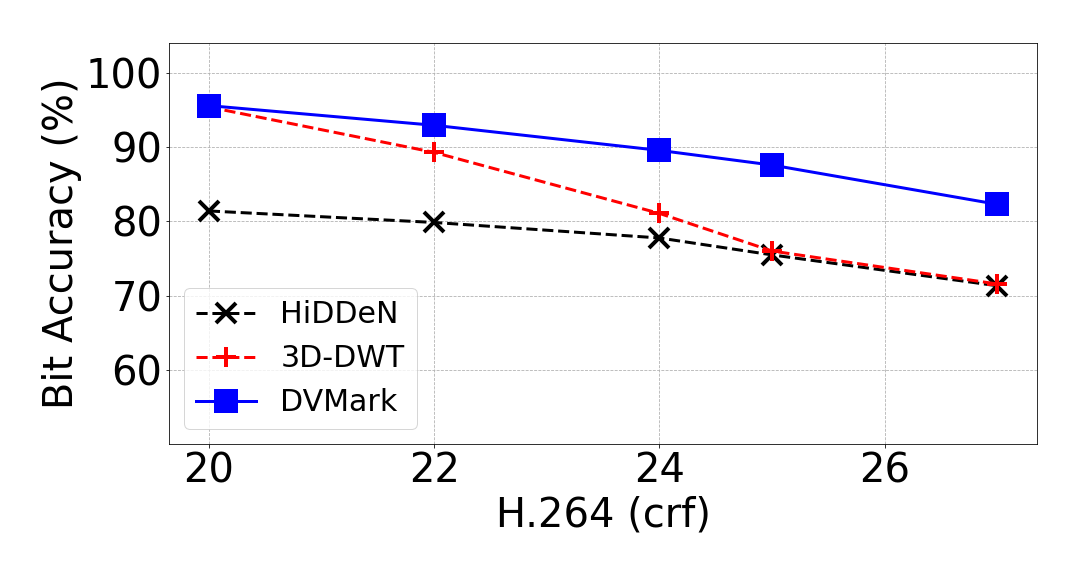}
    \end{tabular}
    \caption{Bit accuracy for 3D-DWT, HiDDeN, and DVMark for various distortions and distortion strengths. For spatial crop, $p$ is the ratio of the cropped box width over the original width, while keeping aspect ratio. For frame drop, $p$ is the proportion of frames randomly dropped from the watermarked video.}
    \vspace{-6mm}
    \label{fig:distortion_strength}
\end{figure*}

As shown in Table~\ref{tab:bit-error-rate}, our method outperforms both the 3D-DWT method as well as HiDDeN by a large margin on almost all tested distortions. We also note that while the 3D-DWT method is fairly robust to some distortions such as blur and Gaussian noise, it is not robust to other types of distortions such as frame drop or crop. This echos with the remark in ~\cite{asikuzzaman2017overview} that it is difficult for existing methods to be robust to a very broad spectrum of distortions. In contrast, our model is simultaneously robust to nearly all of the distortions in Table~\ref{tab:bit-error-rate}, making it a competitive method for practical applications. Figure~\ref{fig:distortion_strength} shows the decoding accuracy when the distortion level is varied. We observe that the gain is consistent across a wide range of strengths. We note that since HiDDeN is applied independently per-frame, it is naturally resistant to frame dropping. However, DVMark managed to match this performance up to a $50\%$ drop rate.
 
Next, we evaluate the quality of the watermarked video with respect to the original in terms of several commonly used full-reference quality metrics, such as Peak signal-to-noise ratio (PSNR)~\cite{hore2010image}, Mean Structural Similarity (MSSIM)~\cite{hore2010image} and LPIPS~\cite{zhang2018unreasonable}. We also report the tLP metric~\cite{chu2020learning}, which is used to specifically measure the temporal consistency of the watermarked videos. In addition to video quality metrics, we also conducted a user study to evaluate the perceptual quality of the watermarked videos. We play a 3-second clip of the encoded video and the watermarked video side-by-side, and ask raters to choose if the pair is ``same",  or ``different". A total of 100 pairs were rated for each watermarking method, where every pair was independently rated by 5 raters. We compute the Mean-Opinion-Score (MOS) as a measure of perceptual quality. From Table~\ref{tab:video-quality}, we observe that our method has superior visual quality compared to HiDDeN and 3D-DWT.

\subsection{Robustness-Quality-Payload Trade-off}
\label{subsec:robustness-quality-payload}
As mentioned in Section~\ref{sec:introduction}, a comprehensive evaluation of a watermarking system requires thorough analysis of the model's payload, robustness, as well as quality. In this section, we visualize this trade-off by fixing either payload or quality, and studying the relationship between the remaining two variables. We use the decoding accuracy as a measure of robustness, PSNR as a measure of quality, and message length for payload. Since the relations also depend on the type of distortion applied to the videos, we select four representative distortions from different categories, namely H.264 compression (CRF=22), spatial cropping (p=0.5), frame drop (p=0.5), and also Gaussian noise (std=0.06), and compute the average decoding accuracy as a measure of robustness.

\begin{table}[!ht]
    \begin{center}
    \begin{adjustbox}{width=0.48\textwidth}
    \begin{tabular}{cccccccc}
    &PSNR $\uparrow$ &MSSIM $\uparrow$ & LPIPS x 100 $\downarrow$ & tLP x 100 $\downarrow$ &MOS $\uparrow$ \\ \hline
    3D-DWT & 36.5& 0.983 & 8.79 & 0.628 & 0.90 \\ \hline
    HiDDeN &35.5 &0.962 & 8.92& 0.354 & 0.86 \\ \hline
    DVMark & \textbf{37.0}& \textbf{0.985}  & \textbf{5.70} & \textbf{0.160} &  \textbf{0.92} \\ \hline    
    \end{tabular}
    \end{adjustbox}
    \end{center}
    \caption{Quality metrics for videos watermarked by 3D-DWT, HiDDeN, and DVMark.}
    \vspace{-5mm}
    \label{tab:video-quality}
\end{table}

\begin{figure}[!ht]
    \centering
    \includegraphics[width=0.95\linewidth]{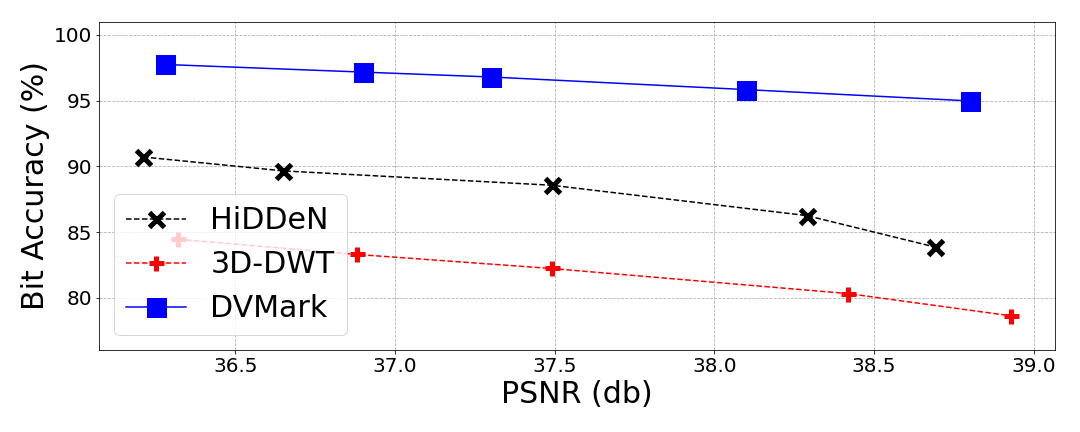}
    \caption{PSNR versus average bit accuracy for DVMark and other methods compared. The curves are generated by adjusting the watermark strength parameter $\alpha$ without re-training the model. }
    \label{fig:psnr_accuracy}
\end{figure}

\begin{figure}[!ht]
    \centering
    \includegraphics[width=0.95\linewidth]{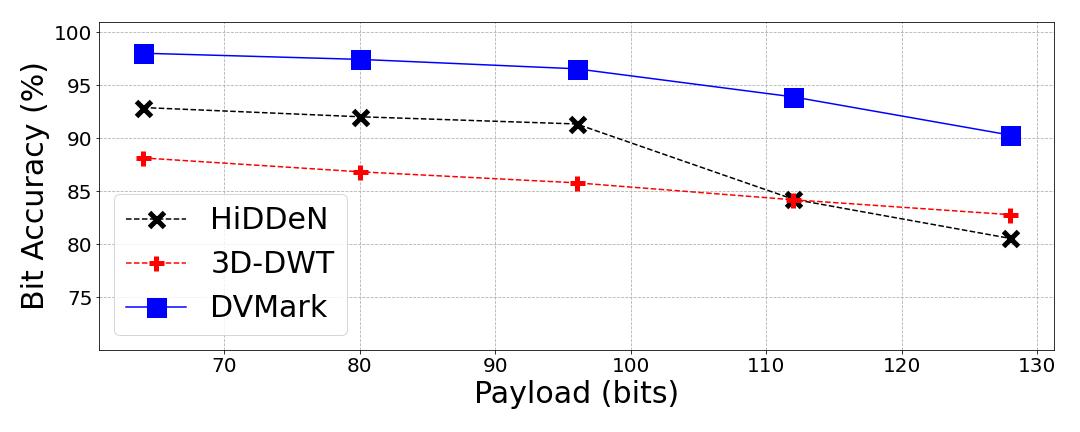}
    \caption{Payload versus average bit accuracy for DVMark and other methods compared.}
    \label{fig:payload_accuracy}
\end{figure}

Figure~\ref{fig:psnr_accuracy} and Figure~\ref{fig:payload_accuracy} shows the trade-off between quality-robustness (for a fixed payload) and payload-robustness (for a fixed quality) respectively. We see that our method is superior to the both 3D-DWT and HiDDeN by a significant margin across both the robustness-quality as well as robustness-payload curves. For Figure~\ref{fig:psnr_accuracy}, we fix the payload to be $m=96$, and adjust the watermark strength $\alpha$ to vary the quality. For Figure~\ref{fig:payload_accuracy}, we train a separate model for each payload, and adjust the watermark strength so that the resulting average PSNR is around 37dB. 

\subsection{Performance on Larger Videos}
A practical watermarking solution must work for a variety of video resolutions and lengths. We demonstrate in this section that our method works out-of-the-box for larger and longer videos. We apply our models fully convolutionally on spatial dimensions, and apply the encoder every $T=8$ frames. Namely, for a video sequence of size $T\times H\times W$, we divide the video to segments of $8 \times H \times W$, and apply DVMark to each of these segments with the same embedded message. We note that in theory our model is fully convolutional in the temporal dimension as well, but we empirically observe that the model performs better with applyting to temporal segments. A possible explanation is due to the limited number of frames $T=8$ at training time. 

\begin{table}[!ht]
    \begin{center}
    \begin{adjustbox}{width=0.48\textwidth}
    \begin{tabular}{cccc}
    Length ($T$) / Resolution ($H\times W$)  & $128 \times 128$ & $462 \times 240$  & $864 \times 480$  \\ \hline
     $T=8$ & 96.14 & 96.63  & 95.29   \\ \hline    
     $T=16$ & 96.60 & 97.07  & 96.00   \\ \hline  
     $T=32$ & 96.80 & 97.16  & 96.01  \\ \hline 
     $T=64$ & 96.80 & 97.19  & 96.18  \\ \hline      

    \end{tabular}
    \end{adjustbox}
    \end{center}
    \caption{Decoding accuracy for DVMark applied to various video resolutions and video lengths. Model accuracy is averaged across four types of distortions (H.264, crop, frame drop, Gaussian noise).}
    \label{tab:accuracy-dimensions}
\end{table}

Same as in Section~\ref{subsec:robustness-quality-payload}, we compute the average decoding accuracy over four distortions (H.264, crop, frame drop, Gaussian noise) for a matrix of different resolutions and video lengths. From the results in Table~\ref{tab:accuracy-dimensions}, we observe no performance degradation when our method is applied to larger videos in general.

\subsection{Detector in a Video Editing Application}
In this section, we evaluate the performance of the watermark detector. The watermark detector head is trained on a mixture of watermarked and unwatermarked video clips. We evaluate the accuracy of the watermark detector on various distortions where we watermark half of the videos at random by the DVMark encoder in our eval set.

\begin{table}[!ht]
    \begin{center}
    \begin{adjustbox}{width=0.48\textwidth}
    \begin{tabular}{ccccccc}
    Identity & \makecell{H.264 \\ (CRF=22)} & \makecell{Frame Average \\(N=3)}  & \makecell{Frame Drop \\ (p=0.5)} &\makecell{Frame Swap \\ (p=0.5)}  \\ \hline
    99.75& 94.27 & 97.10 & 99.75 & 98.89 \\ \hline \hline
    \makecell{Gaussian Blur \\ (2.0)} & \makecell{Gaussian Noise\\ (0.04)} & \makecell{Random Crop \\(p=0.4)}  & \makecell{Random Hue \\ (1.0)} &\makecell{\textbf{Average}} \\ \hline 
     99.50& 97.76 & 98.76 & 99.10 & 98.32 \\ \hline
    \end{tabular}
    \end{adjustbox}
    \end{center}
    \caption{Accuracy of the watermark detector under various video distortions.}
    \vspace{-10pt}
    \label{tab:detector-accuracy}
\end{table}
Table~\ref{tab:detector-accuracy} shows the prediction accuracy of our trained detector. We observe that the detector is able to differentiate between watermarked and unwatermarked video sequences with high accuracy despite the fact that the watermarked videos have good perceptual quality.

\textbf{Video Editing Application}
Next we demonstrate the utility of the watermark detector through a video-editing application. Many modern videos (such as commentary videos) include the use of other original video sources during its creation. In the case that these original video sources are watermarked, it is important to identify and decode the watermarked message even in the presence of unwatermarked content. 

We mimic the video editing pipeline of a commentary video by adding less than 1 second (16 frames) of watermarked source content to an unwatermarked background video of size $462 \times 240$ and of much longer lengths (e.g., T=720 frames). To simulate editing operations to the watermarked content, we center crop the watermarked video to $128\times 128$ and apply a random saturation change to simulate the intermediate editing done by video editors. We finally place the edited source to the lower left corner of the cover video as shown in Figure~\ref{fig:teaser}.  For the most extreme case in our evaluation, the watermarked content consists only $0.5\%$ of the total number of pixels in the final video.

\begin{figure}[!ht]
    \centering
    \includegraphics[ width=0.98\linewidth]{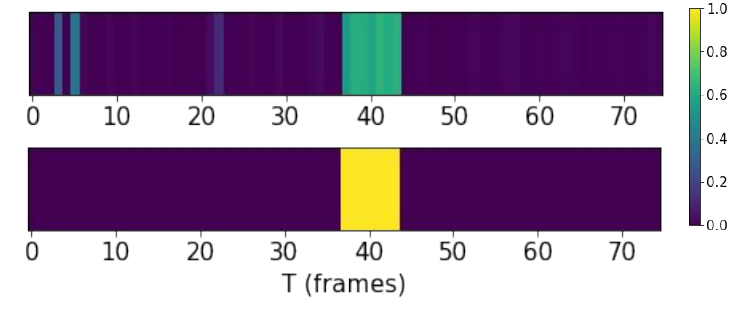}
    \vspace{2mm}
    \caption{Visualization of the watermark detector predictions versus the ground truth. Top row: the probablity predicted by the watmermark detector. Bottom row: The ground truth labels for each frame, 1 indicates that a frame has been watermarked.}
    \label{fig:video-editing-demo}
    \vspace{-3mm}
\end{figure}

We apply the watermark detector to filter the background frames, where we remove frames with a confidence score less than $0.3$ prior to passing the frames to the decoder. We see in Figure~\ref{fig:video-editing-demo} that the detector is able to localize the watermarked video segment with great precision. This greatly improves the ability to decode the message compared to if we simply passed the full video sequence to the decoder, as shown in Table~\ref{tab:video-editing-accuracy}.

\begin{table}[!ht]
    \begin{center}
    \begin{adjustbox}{width=0.48\textwidth}
    \begin{tabular}{ccccccc}
    \hline 
    Background Length & T=60 & T=120  & T=240 & T=360 & T=720 \\ \hline
    With detector & 99.62 & 98.81 & 98.50 & 95.32 & 90.02\\ \hline
    Without detector & 85.19 & 75.31 & 68.40 & 51.08 & 52.14\\ \hline 
    \end{tabular}
    \end{adjustbox}
    \end{center}
    \caption{The decoding accuracy of DVMark for the video editing application with and without detector pre-filtering, for different background video lengths. The background video is unwatermarked, and a watermarked video of length equal 16 frames is inserted to a random interval in the background video.}
    \label{tab:video-editing-accuracy}
    \vspace{-10pt}
\end{table}

\section{Conclusion}
In this work, we propose DVMark, an end-to-end trainable framework for video watermarking. Our method consists of a novel multiscale design where the message is distributed across multiple spatial-temporal scales. Compared to both deep image watermarking and traditional watermarking methods, our method is more robust on a wide range of challenging video distortions, while maintaining good perceptual quality. Moreover, our method easily adapts to other distortion requirements through adding an additional distortion in our end-to-end framework. We further augment our framework with a watermark detector to locate watermarked video segments among unwatermarked content. Through rigorous evaluations, we demonstrate that our DVMark model serves as a practical and reliable solution for video watermarking.

{\small
\bibliographystyle{ieee_fullname}
\bibliography{egbib}
}

\newpage
\section{Additional Design and Experiment Details}

\subsection{Training Details}
\label{sec:training_parameters}
\textbf{Distortions Layer:} We apply a total of 10 differentiable distortions during training. The exact distortions include: CompressionNet (crf=25), random frame drop ($p=0.5$), random frame swap ($p=0.5$), random spatial crop ($p=0.5$), random saturation change, random hue change, 3D-Gaussian blur ($\sigma=2.0$), Gaussian Noise ($\text{std}=0.05$), Differentiable JPEG ($q=50$), and random frame shift, which is a random cyclic permutation of the temporal order. Each distortion is selected with equal probability during training.

\textbf{Encoder-Decoder:}
We train all models with a batch size of 6 and video size $8\times128\times128$, using the ADAM~\cite{kingma2014adam} optimizer with an initial learning rate of 1e-4, and use an exponential decay with decay rate 0.8 every 200k steps. We train for a total of 3M steps. For the loss weights, we set the message loss weight $c_1=0.5$, the GAN generator loss weight $c_2=4e-3$. For training the GAN discriminator, we use the same learning rate but run $N=2$ steps of the discriminator per every generator step. The watermark strength is set to $\alpha = 1.0$ during training.

To avoid spatial and temporal boundary artifacts from encoded videos, we symmetrically pad all of the convolution operations instead of zero-padding. We observe that zero or constant padding results in frames near the temporal boundary (e.g., frame 0 or frame 7) to be visually different from frames in the center, making the model less temporally stable.

\textbf{Detector:}
As described in Section 3.2 of the main text, we train a detector to detect whether a segment of video frames are watermarked or not. The detector head consists of 4 Conv2D operations on top of the feature map $D_f$ followed by a global pooling operation. The logit tensor prior to global pooling can also be used to provide more spatial and temporal granularity. Same as the encoder and decoder, the detector is trained on clips of size $8\times128\times128$ using the ADAM~\cite{kingma2014adam} optimizer with an initial learning rate of 1e-4, and use an exponential decay with decay rate 0.8 every 200k steps. We train for a total of 1M steps. 

\subsection{User Study Details}
For the user study, raters are given the watermarked and original video in a side-by-side fashion, where raters are asked to choose whether the pair is ``identical", or ``different". Raters are encouraged to find difference in terms of noise, temporal flickering, saturation and color changes, as well as other types of differences. Raters are allowed to view the clips as many times as needed. The rating scale is defined as follows: the pair is ``identical" if the rater is unable to find any difference given repeated viewings, ``different" if the rater was able to spot some difference between the unwatermarked and watermarked videos.

\label{sec:architecture}
\subsection{WeightNet Architecture}
In Section 3.2 of the main text, we presented WeightNet, a network that takes the cover video as input and outputs a normalized weight matrix $W$. WeightNet consists of three 3D-Conv layers, with number of channels $32,64,128$ with stride $2$. The output is then globally averaged along the spatial and temporal dimensions, followed by a fully connected layer with $m\times2$ number of units, with $m=96$. Finally a softmax operation is applied such that the output $W_{ij}\in\mathbb{R}^{m\times2}$ is row normalized. 

\subsection{Video GAN Architecture}
Our video GAN discriminator consists of four sub-networks named $\text{ResBlock}_i$, one for each input scale. Each of the sub-networks consists of four residual 3D convolution layers, passed through a global pooling operation on $T, W, H$, and finally passed to a fully connected layer with output channel equal to $1$. The residual layer is defined as follows. Given an input tensor $X$, the residual layer is defined as $R^{(c,s,t)}(X) = \text{conv}^{(c, s, t)}_{3\times3\times3}(X) + \text{conv}^{(c, s, t)}_{1\times1\times1}(X)$, $\text{conv}^{(c, s, t)}$ denotes a 3D convolution with stride $(s,s,t)$ and number of output channels $c$.   The output channel, spatial and temporal stride of the residual layers for each $\text{ResBlock}_i$ is given in Table.~\ref{tab:discriminator architecture}. 

\begin{table}[!htb]
\centering
\begin{adjustbox}{width=0.48\textwidth}
\begin{tabular}{c | c | c | c}
Block  & Channels &  Spatial Strides & Temporal Strides\\
\hline
 $\text{ResBlock}_1$ & (32, 64, 128, 256) & (2, 2, 2, 2) & (1, 1, 1, 1)\\
 \hline
 $\text{ResBlock}_2$ & (32, 64, 128, 256) & (2, 2, 2, 2) & (2, 1, 1, 1)\\
 \hline
 $\text{ResBlock}_3$ & (32, 64, 128, 256) & (2, 2, 2, 1) & (2, 2, 1, 1)\\
 \hline
 $\text{ResBlock}_4$ & (32, 64, 128, 256) & (2, 2, 1, 1) & (2, 2, 2, 1) \\
 \hline 
\end{tabular}
\end{adjustbox}
\vspace{2mm}
\caption{Architecture parameters of the video GAN discriminator.}
\label{tab:discriminator architecture}
\end{table}

\subsection{CompressionNet Architecture}
\begin{figure}[!ht]
    \centering
    \includegraphics[ width=0.98\linewidth]{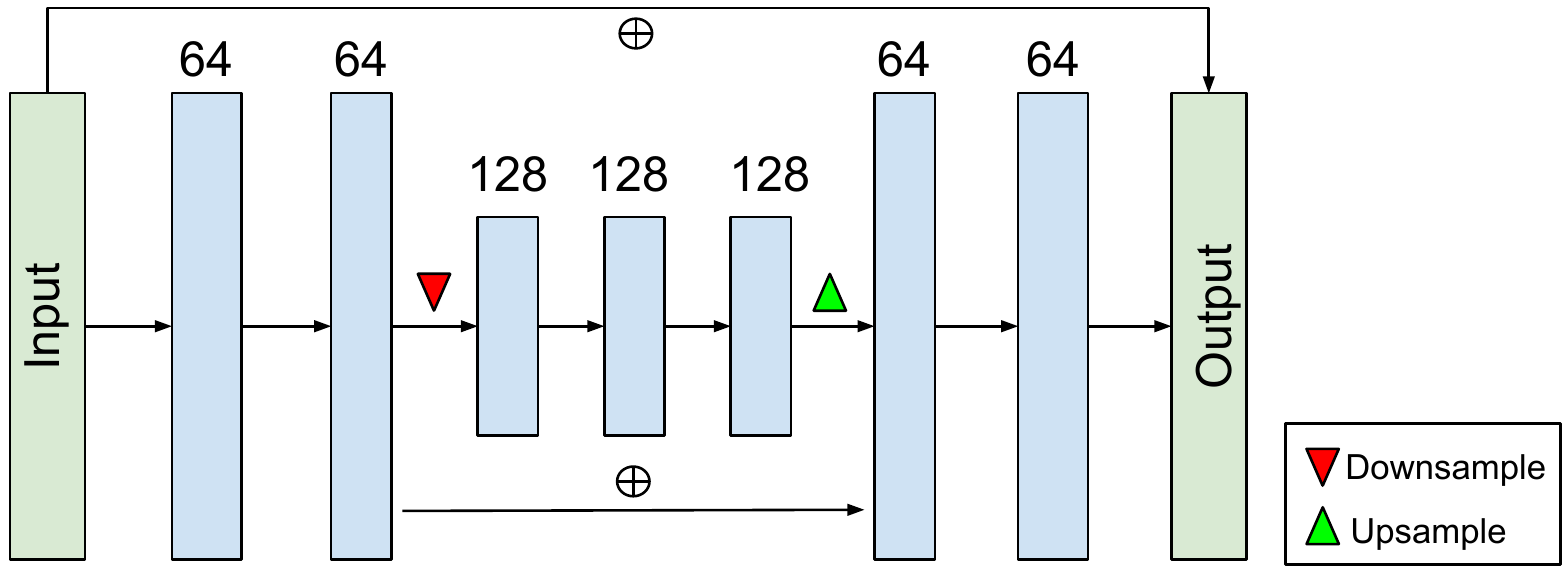} 
    \vspace{2mm}
    \caption{CompressionNet architecture. The numbers on top of each block are the output channels for each layer.}
    \label{fig:compress_net}
\end{figure}

In Section 3.2 of the main text, we presented CompressionNet as a differentiable proxy for video compression distortions. CompressionNet consists of seven layers of 3D convolutions. Each convolution layer has a kernel size of 3, and the number of units for each layer is shown in Figure~\ref{fig:compress_net}. A residual connection is added from the input to the last layer.
We train CompressionNet to mimic the output of an H.264 codec at a fixed Constant Rate Factor (CRF) equal to 25. Same as in DVMark, we train CompressionNet on clips of size $8\times128\times128$.

\section{Additional Detailed Results on Robustness, Quality, Payload}

\subsection{Robustness-Quality Tradeoff}

\begin{table}[!ht]
    \begin{center}
    \begin{adjustbox}{width=0.48\textwidth}
    \begin{tabular}{c|c|c|c|c|c}
     PSNR & 36.21 &36.65  & 37.49 &38.29  & 38.69\\ \hline    
     \makecell{Acc. H.264 (CRF=22)} & 77.90 & 75.11 & 72.21 & 67.92 & 66.71 \\ \hline  
     \makecell{Acc. Crop (p=0.5)}  & 95.48 & 95.16 & 94.14 & 92.92 & 89.31 \\ \hline 
     \makecell{Acc. Frame drop (p=0.5)}  & 99.08 & 99.03 & 98.99 & 97.67 & 96.45  \\ \hline      
     \makecell{Acc. Gaussian noise (std=0.04)}  & 90.37 & 89.29 & 88.89 & 86.50 & 83.03 \\ \hline        
     \makecell{Average}  & 90.71 & 89.65 & 88.56 & 86.25 & 83.87 \\ \hline 
    \end{tabular}
    \end{adjustbox}
    \end{center}
    \caption{Decoding accuracy of HiDDeN for four types of distortions for a range of PSNR values.}
    \label{tab:acc-hidden-quality}
\end{table}

\begin{table}[!ht]
    \begin{center}
    \begin{adjustbox}{width=0.48\textwidth}
    \begin{tabular}{c|c|c|c|c|c}
     PSNR & 36.28 & 36.90 & 37.33 & 38.12 & 38.81\\ \hline    
     \makecell{Acc. H.264 (CRF=22)} & 93.74 & 92.98 & 92.54 & 89.49 & 87.02 \\ \hline  
     \makecell{Acc. Crop (p=0.5)}   & 97.79 & 97.52 & 97.06 & 96.97 & 96.48  \\ \hline 
     \makecell{Acc. Frame drop (p=0.5)}  & 99.17 & 99.15 & 98.98 & 98.96 & 98.91 \\ \hline 
     \makecell{Acc. Gaussian noise (std=0.04)}  & 99.18 & 98.99 & 98.56 & 97.91 & 97.46 \\ \hline     
     \makecell{Average}  & 97.74 & 97.16 & 96.79 & 95.83 & 94.97 \\ \hline 
    \end{tabular}
    \end{adjustbox}
    \end{center}
    \caption{Decoding accuracy of 3D-DWT for four types of distortions for a range of PSNR values.}
    \label{tab:acc-dwt-quality}
\end{table}

\begin{table}[!ht]
    \begin{center}
    \begin{adjustbox}{width=0.48\textwidth}
    \begin{tabular}{c|c|c|c|c|c}
     PSNR & 36.32 &36.88  & 37.49 &38.42  & 38.93\\ \hline    
     \makecell{Acc. H.264 (CRF=22)} &  86.48 & 82.91 & 79.46 & 73.17 & 68.06\\ \hline  
     \makecell{Acc. Crop (p=0.5)}  & 70.75 & 71.08 & 70.88 & 71.07 & 70.75\\ \hline 
     \makecell{Acc. Frame drop (p=0.5)}  & 81.02 & 79.71 & 79.15 & 77.84 & 76.89\\ \hline      
     \makecell{Acc. Gaussian noise (std=0.04)}  & 99.54 & 99.49 & 99.41 & 99.18 & 98.82\\ \hline      
     \makecell{Average}  & 84.45 & 83.30 & 82.23 & 80.32 & 78.63\\ \hline 
    \end{tabular}
    \end{adjustbox}
    \end{center}
    \caption{Decoding accuracy of DVMark for four types of distortions for a range of PSNR values.}
    \label{tab:acc-dvmark-quality}
\end{table}

In Tables~\ref{tab:acc-hidden-quality} to~\ref{tab:acc-dvmark-quality}, we provide the exact PSNR values and decoding accuracy values for four distortions (H.264, Crop, Frame drop, Gaussian noise) used to plot Figure 7 in the main text. While Figure 7 provides an overview of the robustness-quality trade-off of DVMark, HiDDeN, and 3D-DWT, we also provide these tables as a reference.

\subsection{Robustness-Payload Tradeoff}

In Tables~\ref{tab:acc-hidden-payload} to~\ref{tab:acc-dvmark-payload}, we provide the decoding accuracy values for four distortions (H.264, Crop, Frame drop, Gaussian noise), as well as the payload used to plot Figure 8 in the main text. While Figure 8 provides an overview of the robustness-payload trade-off of DVMark, HiDDeN, and 3D-DWT, we also provide these tables as a reference.

\begin{table}[!ht]
    \begin{center}
    \begin{adjustbox}{width=0.48\textwidth}
    \begin{tabular}{c|c|c|c|c|c}
     Payload (bits) & 64 & 80 & 96 & 112 & 128\\ \hline    
     \makecell{Acc. H.264 (CRF=22)} & 80.44  & 79.61 & 79.85 & 69.23 & 60.97 \\ \hline  
     \makecell{Acc. Crop (p=0.5)}  & 96.51 & 95.85 & 95.27 & 90.23 & 89.91\\ \hline 
     \makecell{Acc. Frame drop (p=0.5)}  & 99.54 & 99.36 & 99.03 & 93.10 & 91.21\\ \hline      
     \makecell{Acc. Gaussian noise (std=0.04)}  & 95.11 & 93.28 & 91.27 & 84.42 & 80.18\\ \hline      
     \makecell{Average}  & 92.9 & 92.03 & 91.35 & 84.25 & 80.56 \\ \hline 
    \end{tabular}
    \end{adjustbox}
    \end{center}
    \caption{Decoding accuracy of HiDDeN for four types of distortions for a range of message lengths (payload).}
    \label{tab:acc-hidden-payload}
\end{table}

\begin{table}[!ht]
    \begin{center}
    \begin{adjustbox}{width=0.48\textwidth}
    \begin{tabular}{c|c|c|c|c|c}
     Payload (bits) & 64 & 80 & 96 & 112 & 128\\ \hline    
     \makecell{Acc. H.264 (CRF=22)} & 93.01 & 92.90 & 92.94 & 89.03 & 84.63\\ \hline  
     \makecell{Acc. Crop (p=0.5)}  & 76.36 & 73.74 & 71.38 & 70.91 & 69.40\\ \hline 
     \makecell{Acc. Frame drop (p=0.5)}  & 83.29 & 81.20 & 79.85 & 77.69 & 78.03\\ \hline      
     \makecell{Acc. Gaussian noise (std=0.04)}  & 99.93 & 99.50 & 99.01 & 99.17 & 99.15 \\ \hline      
     \makecell{Average}  & 88.15 & 86.84 & 85.80 & 84.2 & 82.80\\ \hline 
    \end{tabular}
    \end{adjustbox}
    \end{center}
    \caption{Decoding accuracy of 3D-DWT for four types of distortions for a range of message lengths (payload).}
    \label{tab:acc-dwt-payload}
\end{table}

\begin{table}[!ht]
    \begin{center}
    \begin{adjustbox}{width=0.48\textwidth}
    \begin{tabular}{c|c|c|c|c|c}
     Payload (bits) & 64 & 80 & 96 & 112 & 128\\ \hline   
     \makecell{Acc. H.264 (CRF=22)} & 93.50 & 93.11 & 92.94 & 88.07 & 83.09\\ \hline  
     \makecell{Acc. Crop (p=0.5)}  & 98.72 & 98.10 & 97.06 & 94.21 & 87.72\\ \hline 
     \makecell{Acc. Frame drop (p=0.5)}  & 99.90 & 99.12 & 97.64 & 97.04 & 96.82\\ \hline      
     \makecell{Acc. Gaussian noise (std=0.04)}  & 99.95 & 99.43 & 98.56 & 96.28 & 93.57\\ \hline      
     \makecell{Average}  & 98.02 & 97.44 & 96.55 & 93.9 & 90.3 \\ \hline    
    \end{tabular}
    \end{adjustbox}
    \end{center}
    \caption{Decoding accuracy of DVMark for four types of distortions for a range of message lengths (payload).}
    \label{tab:acc-dvmark-payload}
\end{table}

\subsection{Performance on Larger Videos}
In Tables~\ref{tab:vary-dimension-h264} to~\ref{tab:vary-dimension-gaussian}, we present the performance of DVMark on various video lengths and resolutions on four distortions (H.264, crop, frame drop, Gaussian noise). The average of these values are already shown in Table 3 in the main text.
\begin{table}[!ht]
    \begin{center}
    \begin{adjustbox}{width=0.48\textwidth}
    \begin{tabular}{cccc}
    Length ($T$) / Resolution ($H\times W$)  & $128 \times 128$ & $462 \times 240$  & $864 \times 480$  \\ \hline
     $T=8$ & 92.94 & 92.25  & 86.15   \\ \hline    
     $T=16$ & 93.02 & 93.18  & 88.26   \\ \hline  
     $T=32$ & 92.98 & 93.20  & 88.32  \\ \hline 
     $T=64$ & 92.96 & 93.19  & 88.73 \\ \hline      

    \end{tabular}
    \end{adjustbox}
    \end{center}
    \caption{Decoding accuracy for DVMark applied to various video resolutions and video lengths with H.264 distortion with CRF=22.}
    \label{tab:vary-dimension-h264}
\end{table}

\begin{table}[!ht]
    \begin{center}
    \begin{adjustbox}{width=0.48\textwidth}
    \begin{tabular}{cccc}
    Length ($T$) / Resolution ($H\times W$)  & $128 \times 128$ & $462 \times 240$  & $864 \times 480$  \\ \hline
     $T=8$ & 96.46 & 98.77  & 99.60   \\ \hline    
     $T=16$ & 97.35 & 98.82 & 99.54   \\ \hline  
     $T=32$ & 97.58 & 98.89  & 99.57  \\ \hline 
     $T=64$ & 97.52 & 98.90  & 99.55  \\ \hline      

    \end{tabular}
    \end{adjustbox}
    \end{center}
    \caption{Decoding accuracy for DVMark applied to various video resolutions and video lengths with spatial crop with crop ratio 0.4.}
    \label{tab:vary-dimension-crop}
\end{table}

\begin{table}[!ht]
    \begin{center}
    \begin{adjustbox}{width=0.48\textwidth}
    \begin{tabular}{cccc}
    Length ($T$) / Resolution ($H\times W$)  & $128 \times 128$ & $462 \times 240$  & $864 \times 480$  \\ \hline
     $T=8$ & 96.59 & 96.95  & 97.01   \\ \hline    
     $T=16$ & 97.51 & 97.64  & 97.75   \\ \hline  
     $T=32$ & 97.91 & 97.98  & 98.01  \\ \hline 
     $T=64$ & 98.02 & 98.09  & 98.02  \\ \hline      

    \end{tabular}
    \end{adjustbox}
    \end{center}
    \caption{Decoding accuracy for DVMark applied to various video resolutions and video lengths with frame drop distortion (p=0.5).}
    \label{tab:vary-dimension-frame-drop}
\end{table}

\begin{table}[!ht]
    \begin{center}
    \begin{adjustbox}{width=0.48\textwidth}
    \begin{tabular}{cccc}
    Length ($T$) / Resolution ($H\times W$)  & $128 \times 128$ & $462 \times 240$  & $864 \times 480$  \\ \hline
     $T=8$ & 98.56 & 98.53  & 98.38   \\ \hline    
     $T=16$ & 98.51 & 98.64  & 98.41  \\ \hline  
     $T=32$ & 98.72 & 98.60  & 98.44  \\ \hline 
     $T=64$ & 98.69 & 98.60  & 98.42  \\ \hline      

    \end{tabular}
    \end{adjustbox}
    \end{center}
    \caption{Decoding accuracy for DVMark applied to various video resolutions and video lengths with Gaussian Noise with stddev=0.04.}
    \label{tab:vary-dimension-gaussian}
\end{table}

We note that the effect of adjusting the video resolution (from the same source video) is a trade-off between more spatial redundancy versus less high-frequency regions in the image. In terms of distortions, the effects of increasing resolution are distortion dependent. For crop and frame drop, increasing spatial resolution in general improves the decoding accuracy, whereas for Gaussian noise, the effect is neutral. For H.264, we see a slight drop for the largest video resolution. For video length, we see that a longer video increases the decoding accuracy for all video resolutions and distortion types.

\section{Ablation Studies}

\subsection{A Study of Per-video Robustness}
The content of a video is certainly an important factor for the performance of any watermarking method, but is rarely discussed in the literature.  We study for which videos the model performs the best and worst in terms of the decoding accuracy.

\begin{figure}
    \centering
    \includegraphics[ width=1.0\linewidth]{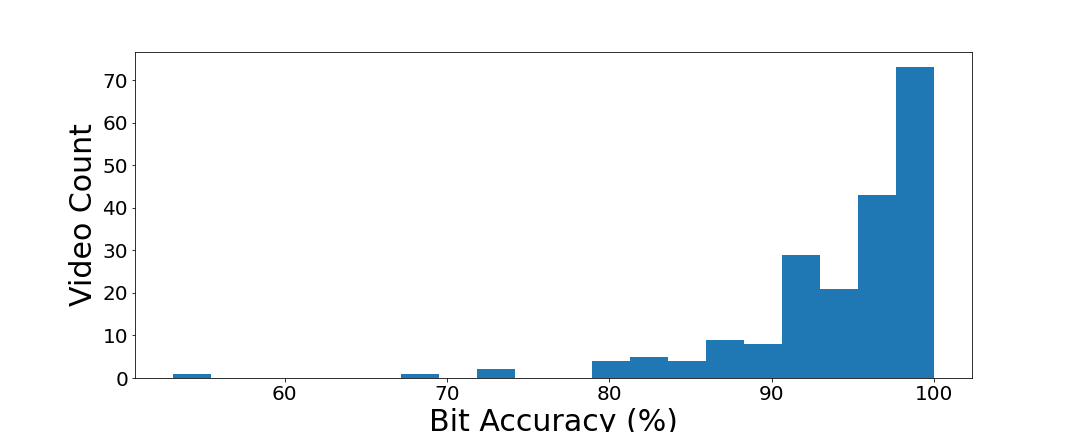}
    \caption{Histogram of bit accuracy collected for 200 video clips of length 8 and resolution $128 \times 128$. We apply the DVMark encoder and distort the video by applying Gaussian noise with a standard deviation of $0.06$. }
    \label{fig:bit_hist}
\end{figure}

\begin{figure}
    \centering
    \includegraphics[ width=1.0\linewidth]{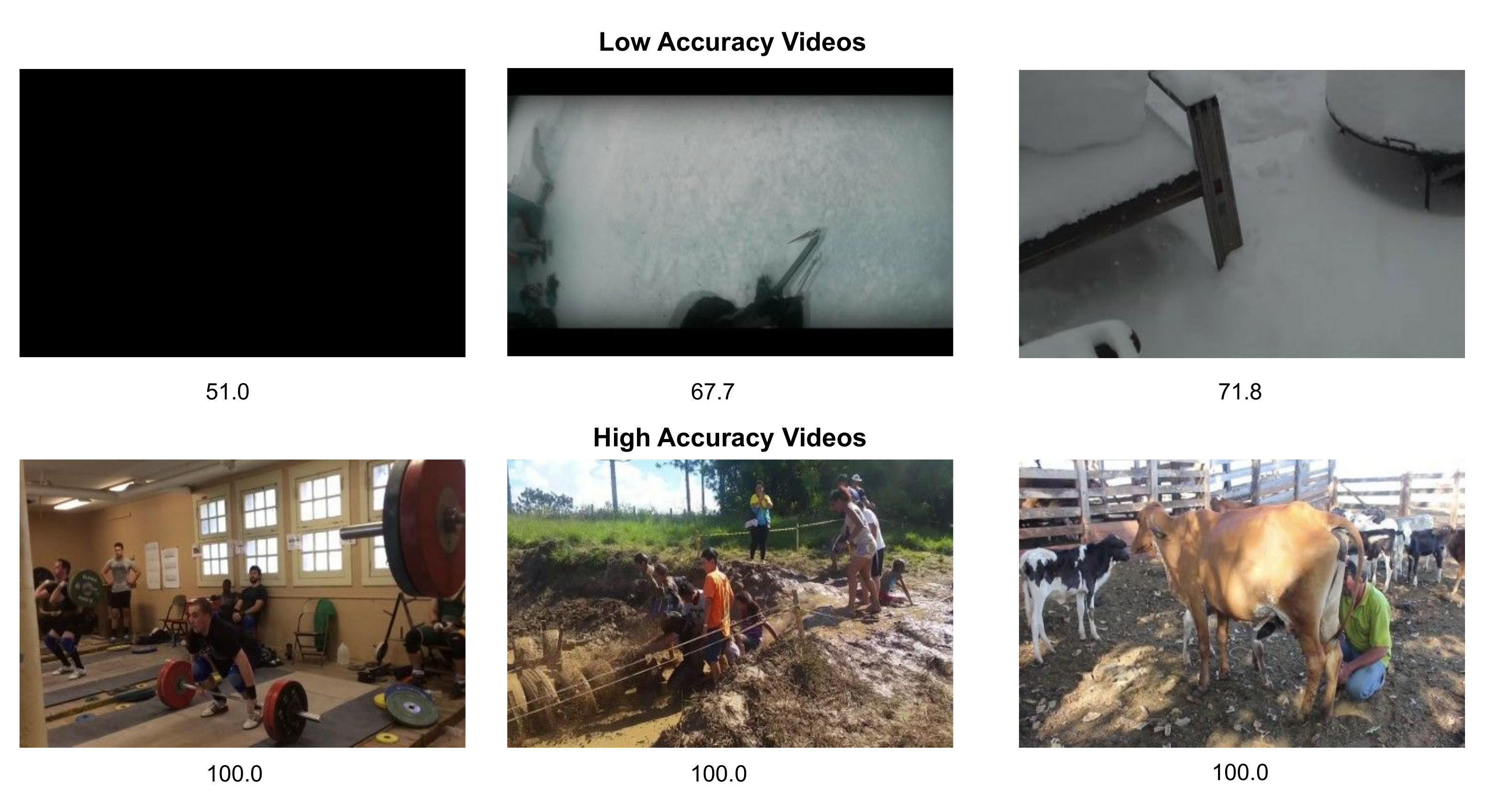}
    \caption{Samples of video clips with low decoding accuracy versus high decoding accuracy. The bit accuracy of the example clip is given in the numbers below each frame.}
    \label{fig:bit_samples}
\end{figure}

Figure~\ref{fig:bit_hist} plots the histogram of decoding accuracy aggregated over 200 videos clips with length equal 8 frames. To avoid accuracy values concentrating at 100.0\%, we apply a the Gaussian Noise distortion with standard deviation $0.06$ to make the decoding more challenging. 

We note that while the majority of the clips have accuracy higher than $90\%$, there are a few examples with relatively low accuracy. We visualize these examples in Figure~\ref{fig:bit_samples}. We observe that videos with less high frequency components (e.g., blurry videos) and with less motion perform worse than videos with more texture and more motion. This aligns well with the intuition that videos with ``richer" content are easier to watermark. 

\subsection{Network Architecture}
In this section, we demonstrate the effectiveness of our multi-head design.  Table~\ref{tab:acc-network} shows the decoding accuracy of various design choices over various distortions. We observe that the multiscale components improve the robustness of the overall method.
\begin{table}[!ht]
    \begin{center}
    \begin{adjustbox}{width=0.48\textwidth}
    \begin{tabular}{c|c|c|c|c}
     Payload (bits) & Head1 & Head2 & Head1 + Head2 & Head1 + Head2 +WeightNet \\ \hline    
     \makecell{Identity} & 99.80  &  99.81 & 99.76 & 99.85 \\ \hline  
     \makecell{H.264 (crf=22) }  &88.26 & 87.74 & 92.85 & 92.94 \\ \hline 
     \makecell{Gaussian Noise (std=0.04)}  & 94.57 & 96.00 &96.79 & 97.07  \\ \hline      
     \makecell{Average}  & 94.21 & 94.52 & 96.47 & 96.62 \\ \hline 
    \end{tabular}
    \end{adjustbox}
    \end{center}
    \caption{Decoding accuracy of several variations of the network design. ``Head1" and ``Head2" refers to the case where one of the decoder heads is removed. ``Head1 + Head2" refers to the case where we use both decoder heads but simply sum the output without weighting.  ``Identity" distortion refers to the case where no distortion is applied.}
    \label{tab:acc-network}
\end{table}

\subsection{Loss Function Weights}
In this section, we investigate how the various weights $c_1$ and $c_2$ in the loss function (Equation 5 in the main text) affect the trained model. The loss function is defined as
\begin{equation}
    L_{total} = L_I(V_{in}, V_{w}) + c_1L_M(M, M^w) + c_2 L_G(V_{w}),
\label{eq:total_loss}
\end{equation}
with $c_1$ being the message loss weights and $c_2$ being the GAN loss.

$c_1$ controls the trade-off between quality and robustness. Table~\ref{tab:acc-message-weight} shows the decoding accuracy and PSNR of the model when $c_1$ is varied. We fix the other hyper-parameters same as in Section 1.1.
\begin{table}[!ht]
    \begin{center}    \begin{adjustbox}{width=0.43\textwidth}
    \begin{tabular}{c|c|c|c|c}
     Message Loss Weight &  $c_1=$0.1 &$c_1=$0.3 &$c_1=$ 0.5 & $c_1=$0.7 \\ \hline    
     \makecell{Accuracy} & 85.3  &  92.2 & 96.6 & 98.7 \\ \hline  
     \makecell{PSNR }  &39.1 & 37.8 & 36.8 & 34.2 \\ \hline 
    \end{tabular} \end{adjustbox}
    \end{center}
    \caption{Decoding accuracy versus PSNR for various message loss weight $c_1$. The accuracy is computed as an average of Identity, H.264, and Gaussian Noise distortions. }
    \label{tab:acc-message-weight}
\end{table}

\begin{figure}
    \centering
    \includegraphics[ width=1.0\linewidth]{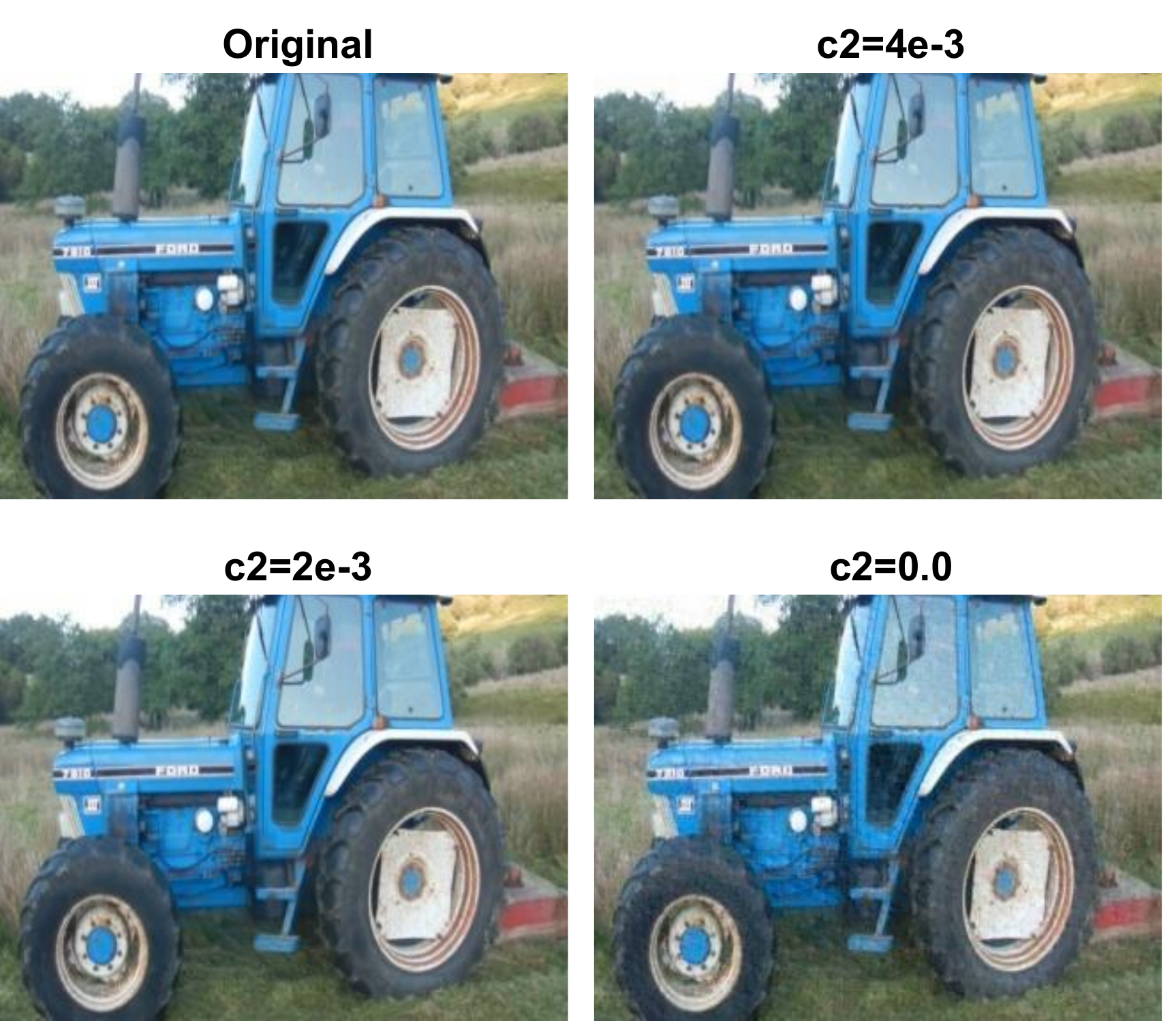}
    \caption{Samples of the original frame and watermarked frames with varying GAN loss weight $c_2$. Note that frames without the GAN loss ($c_2=0.0$) are much noisier visually. }
    \label{fig:gan_samples}
\end{figure}

$c_2$ controls the scale of the perceptual quality versus $L_2$ loss. Figure~\ref{fig:gan_samples} shows that lower GAN weight tends to generate noisier watermarks that are less aesthetically pleasing.

\subsection{CompressionNet}

To evaluate the effectiveness of various approaches for improving robustness to video compression, we compare the performance of CompressionNet with several other alternative methods that dampen the high-frequency components of a video. The methods selected are 3D and 2D Gaussian blurring, frequency truncation, and differentiable JPEG~\cite{shin2017jpeg}. For 2D and 3D Gaussian blur, we use a Gaussian kernel with spatial kernel size $5$ and standard deviation $2.0$. We use a temporal kernel size $3$ for 3D Gaussian blur. For differentiable JPEG, we set a quality factor of $q=50$ with chroma-subsampling enabled. For frequency truncation, we apply a 3D-FFT and set the coefficient outside a cube of width $0.5T \times 0.5H \times 0.5W$ to $0$. Table~\ref{tab:video-distortion-performance} shows that while these methods are effective to some extent at providing robustness against compression, they are less effective compared to CompressionNet.

\begin{table}[!ht]
    \begin{center}
    \begin{adjustbox}{width=0.48\textwidth}
    \begin{tabular}{ccccccc}
    \hline
     & \makecell{3D-Blur} & \makecell{2D-Blur} & \makecell{Diff-\\JPEG} & \makecell{Identity} & \makecell{Compression-\\Net} & \makecell{Combined} \\ \hline
    CRF=22  & 78.1& 75.6& 72.4 & 53.5  & 89.3 & 93.0\\ \hline
    \end{tabular}
    \end{adjustbox}    
    \end{center}
    \caption{Bit accuracy on H.264 video compression for models trained with different types of differentiable distortions.}
    \label{tab:video-distortion-performance}
    \vspace{-3mm}
\end{table}

\section{Visual Examples}
\label{sec:examples}

\begin{figure*}[!ht]
    \centering
    \begin{tabular}{c}
    \hspace{-0.25cm}
    \includegraphics[width=0.99\linewidth]{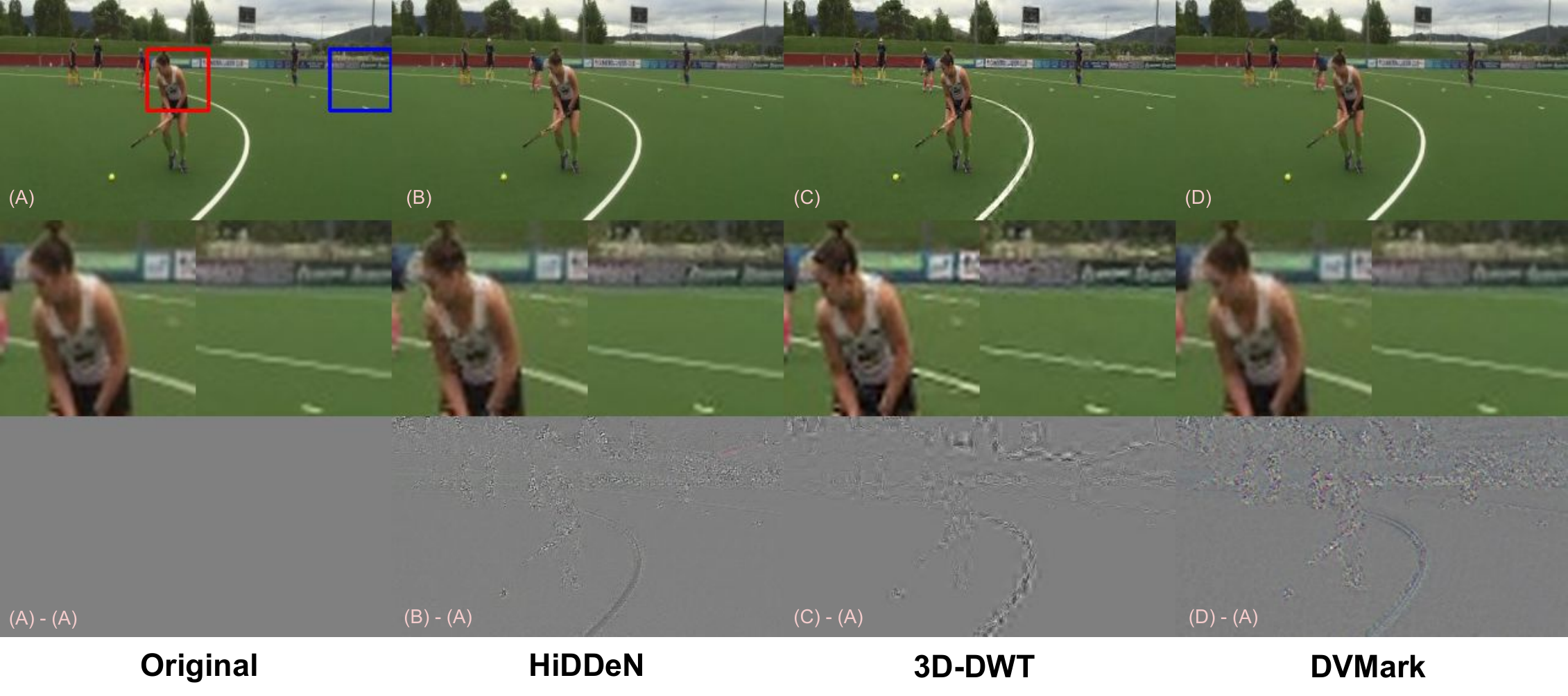}
    \end{tabular}
    \caption{Visual samples 1 of original and watermarked video frames from various watermarking methods.  The second row contains a zoomed in view to better display the details.  Note the ``Difference" image (third row) is the \emph{scaled} to 0-1 for each example, and thus the magnitude does not reflect the absolute value of the difference. }
    \vspace{-5mm}
    \label{fig:visual_samples-1}
\end{figure*}

More visual samples from DVMark, as well as HiDDeN and 3D-DWT can be found in fsup~\ref{fig:visual_samples-1} to~\ref{fig:visual_samples-5}. Note that the ``Difference" image (third row) is the \emph{scaled} difference (normalized to 0-1) between the watermarked and cover frames. Thus the magnitude does not reflect the absolute value of the difference. The difference map serves only as a visualization of where the information is embedded in the signal.

\begin{figure*}[!ht]
    \centering
    \begin{tabular}{c}
    \hspace{-0.25cm}
    \includegraphics[width=0.99\linewidth]{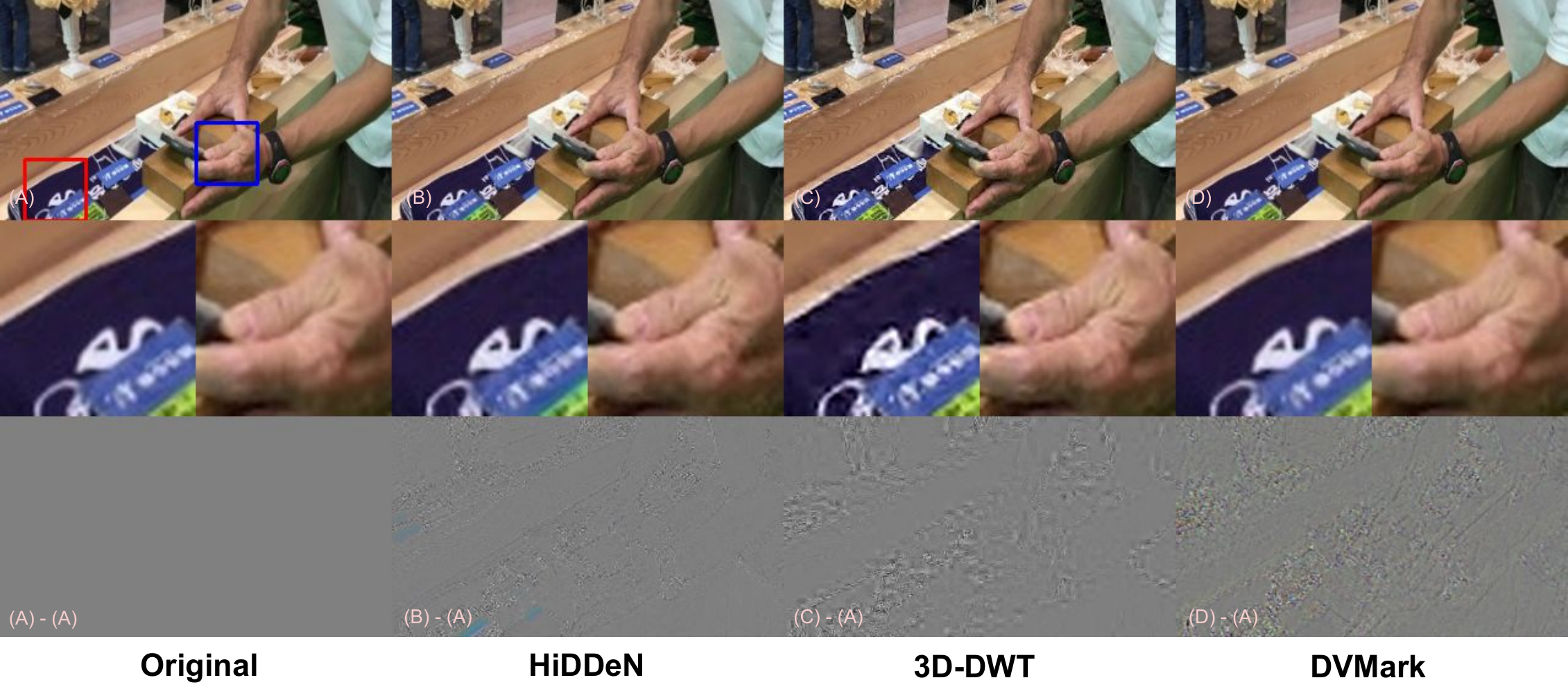}
    \end{tabular}
    \caption{Visual samples 2 of original and watermarked video frames from various watermarking methods. The second row contains a zoomed in view to better display the details.  Note the ``Difference" image (third row) is the \emph{scaled} to 0-1 for each example, and thus the magnitude does not reflect the absolute value of the difference. }
    \vspace{-5mm}
    \label{fig:visual_samples-2}
\end{figure*}

\begin{figure*}[!ht]
    \centering
    \begin{tabular}{c}
    \hspace{-0.25cm}
    \includegraphics[width=0.99\linewidth]{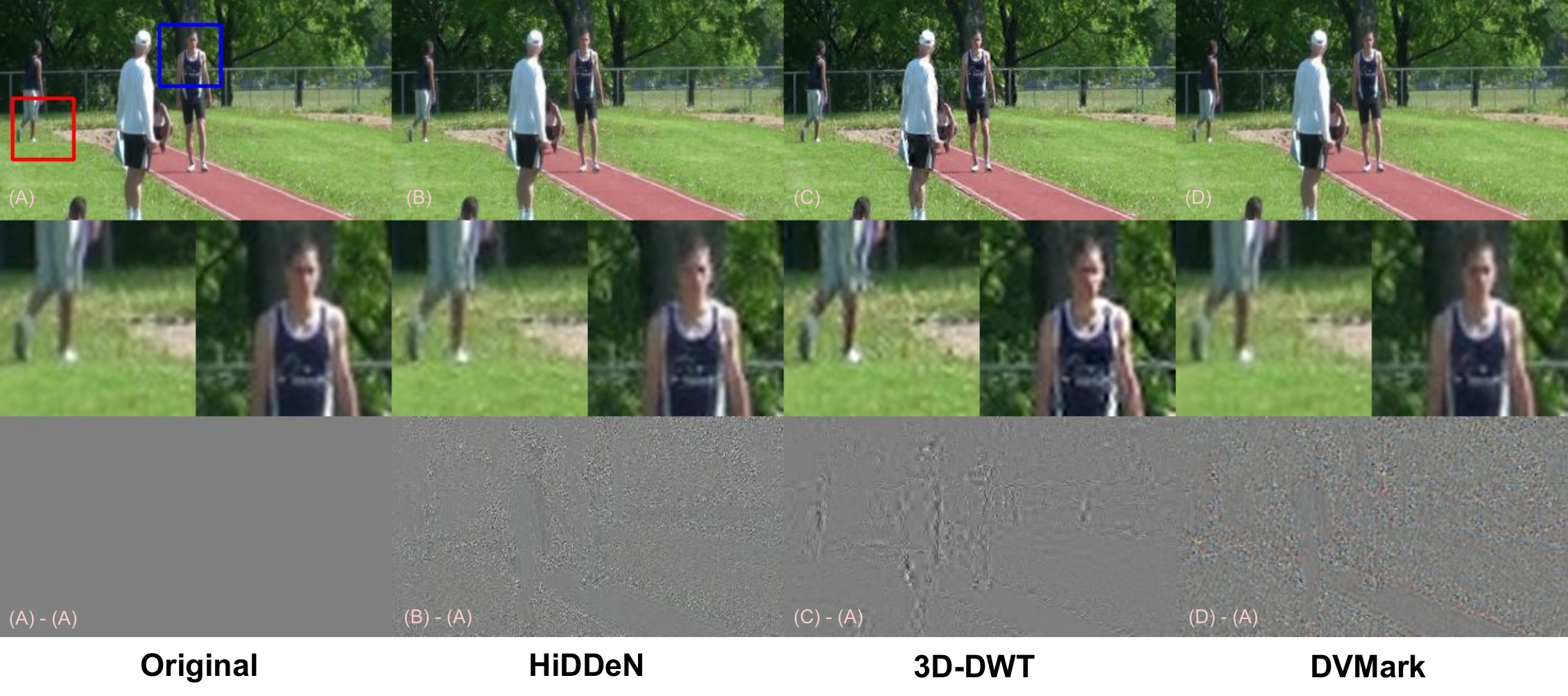}
    \end{tabular}
    \caption{Visual samples 3 of original and watermarked video frames from various watermarking methods. The second row contains a zoomed in view to better display the details.  Note the ``Difference" image (third row) is the \emph{scaled} to 0-1 for each example, and thus the magnitude does not reflect the absolute value of the difference. }
    \vspace{-5mm}
    \label{fig:visual_samples-3}
\end{figure*}

\begin{figure*}[!ht]
    \centering
    \begin{tabular}{c}
    \hspace{-0.25cm}
    \includegraphics[width=0.99\linewidth]{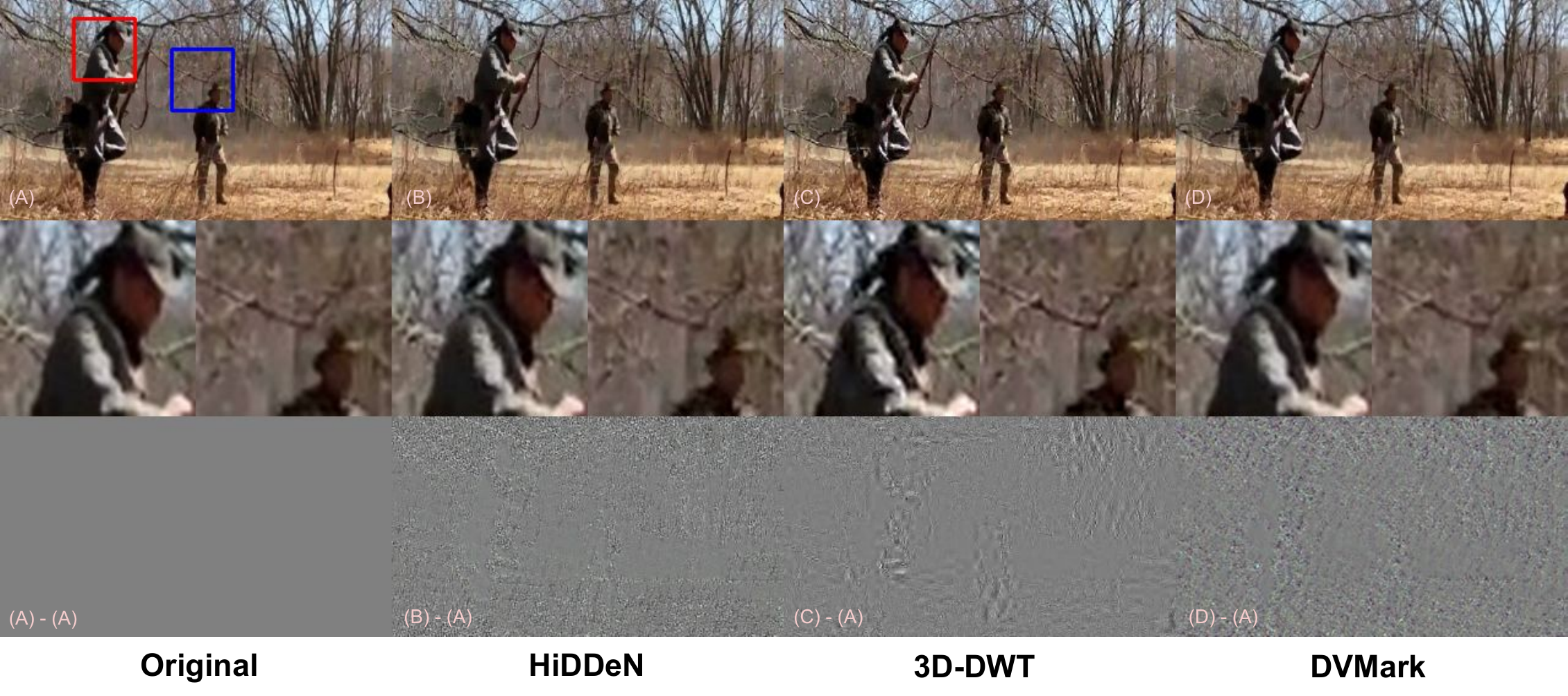}
    \end{tabular}
    \caption{Visual samples 4 of original and watermarked video frames from various watermarking methods. The second row contains a zoomed in view to better display the details.  Note the ``Difference" image (third row) is the \emph{scaled} to 0-1 for each example, and thus the magnitude does not reflect the absolute value of the difference. }
    \vspace{-5mm}
    \label{fig:visual_samples-4}
\end{figure*}

\begin{figure*}[!ht]
    \centering
    \begin{tabular}{c}
    \hspace{-0.25cm}
    \includegraphics[width=0.99\linewidth]{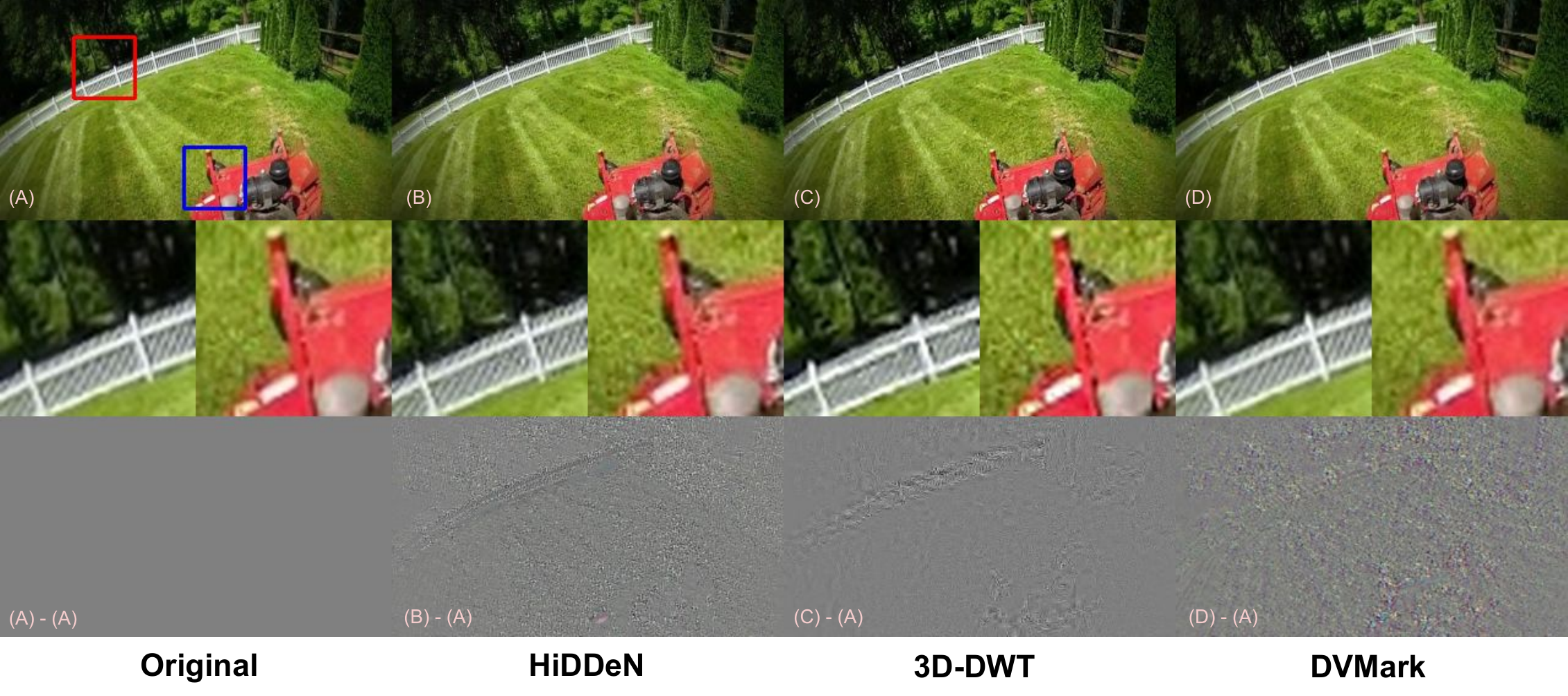}
    \end{tabular}
    \caption{Visual samples 5 of original and watermarked video frames from various watermarking methods. The second row contains a zoomed in view to better display the details.  Note the ``Difference" image (third row) is the \emph{scaled} to 0-1 for each example, and thus the magnitude does not reflect the absolute value of the difference. }
    \vspace{-5mm}
    \label{fig:visual_samples-5}
\end{figure*}

\begin{figure*}[!ht]
    \centering
    \begin{tabular}{c}
    \hspace{-0.25cm}
    \includegraphics[width=0.99\linewidth]{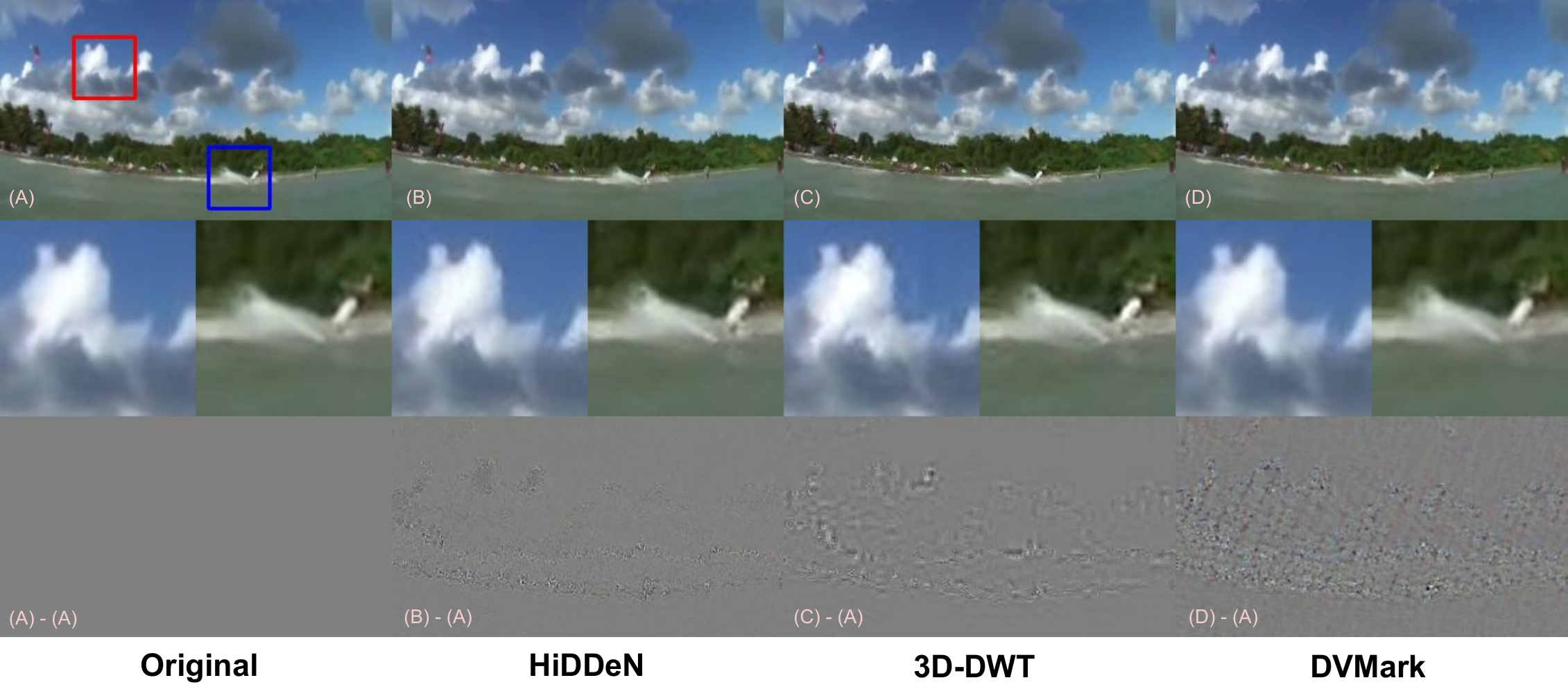}
    \end{tabular}
    \caption{Visual samples 6 of original and watermarked video frames from various watermarking methods. The second row contains a zoomed in view to better display the details. Note the ``Difference" image (third row) is the \emph{scaled} to 0-1 for each example, and thus the magnitude does not reflect the absolute value of the difference. }
    \vspace{-5mm}
    \label{fig:visual_samples-5}
\end{figure*}

\end{document}